\documentclass[aps, prc, twocolumn, tightenlines, letterpaper, amsmath, amssymb, preprintnumbers, superscriptaddress, floatfix, longbibliography, nofootinbib]{revtex4-2}

\usepackage[T1]{fontenc}
\usepackage{bm}
\usepackage{xspace}
\usepackage[dvipsnames]{xcolor}
\usepackage{graphicx}
\usepackage{tabularx}
\usepackage{enumitem}
\usepackage{braket}
\usepackage{dsfont}
\usepackage[normalem]{ulem}

\usepackage{amsmath, nccmath}
\usepackage{lipsum}
\usepackage{cancel}
\usepackage{bm}
\usepackage[linesnumbered,ruled,vlined]{algorithm2e}
\usepackage{algpseudocode}
\usepackage{physics}
\usepackage{amsfonts}

\newcolumntype{Y}{>{\centering\arraybackslash}X}
\usepackage{diagbox}

\DeclareMathAlphabet{\mathpzc}{OT1}{pzc}{m}{it}

\usepackage[hypertexnames=false]{hyperref}
\hypersetup{
    colorlinks=true,       % false: boxed links; true: colored links
    linkcolor=blue,          % color of internal links
    citecolor=blue,        % color of links to bibliography
    filecolor=blue,      % color of file links
    urlcolor=blue           % color of external links
}

\usepackage{orcidlink}

\newcommand{\vbar}{\bar{v}}
\newcommand{\gbar}{\bar{g}}
\newcommand{\M}{\mathcal{M}}

\usepackage{mathtools} %for coloneqq
\usepackage[caption=false]{subfig}

\begin{document}

\title{Quantum Complexity and Stabilizer-State Expansions: From Pairing to Deformation in Fermionic Systems}

\author{S. Momme Hengstenberg\,\orcidlink{0009-0009-8764-9980}}
\email{shengstenberg@physik.uni-bielefeld.de}
\affiliation{Fakult\"at f\"ur Physik, Universit\"at Bielefeld, D-33615, Bielefeld, Germany}

\author{Caroline E. P. Robin\,\orcidlink{0000-0001-5487-270X}}
\email{crobin@physik.uni-bielefeld.de}
\affiliation{Fakult\"at f\"ur Physik, Universit\"at Bielefeld, D-33615, Bielefeld, Germany}
\affiliation{GSI Helmholtzzentrum f\"ur Schwerionenforschung, Planckstra{\ss}e 1, 64291 Darmstadt, Germany}

\begin{abstract}
Understanding how collective phenomena such as pairing and deformation emerge in interacting fermionic systems, and how the correlations underlying them shape the resources required for their simulation, is a central challenge in many-body physics.
We investigate various aspects of quantum complexity in a fermionic system exhibiting two main forms of collectivity, specifically deformation and pairing, through entanglement, non-stabilizerness, and fermionic non-Gaussianity. In the deformed phase, high entanglement coexists with low non-stabilizerness, while the superfluid pairing phase, by contrast, exhibits substantial multipartite entanglement and non-stabilizerness, together with maximal fermionic anti-flatness.
In a second step we examine how to describe the phase diagram of the system in a classically scalable and symmetry-preserving way, by combining stabilizer-state expansion and variational basis optimization.
A physics-guided decomposition of the Hamiltonian identifies the relevant stabilizer states, which can be constructed from graph states.
A single stabilizer state captures the deformed phase to a large extent, and,
although the number of stabilizer states required to describe the pairing phase grows exponentially with system size, a symmetry-preserving basis rotation yields a high-fidelity approximation using far fewer states. This rotated state can then serve as a practical starting point for quantum refinement.
\end{abstract}

\maketitle

%%%%%%%%%%%%%%%%%%%%%%%%%%%%%%%%
\section{Introduction}
\noindent

Describing accurately the structure and dynamics of interacting quantum many-body (QMB) systems and their emergent behaviors is of paramount importance for fundamental physics and applied sciences, including computing, sensing, as well as materials and drug discoveries. 
It is now well recognized that achieving predictive simulations of such systems will require integrated classical-quantum computing ecosystems, where computational tasks are strategically distributed, and quantum processors are invoked for tasks that are beyond the efficient reach of classical methods.
The optimal balance between classical and quantum resources is guided by the underlying physical properties of the target systems, and continuously evolves as developments in quantum information also drive new insights and improvements in classical methods. 

Making connections between physical phenomena and aspects of quantum complexity, such as entanglement, non-stabilizerness, or non-gaussianity, which are in direct relation with computational resources, may be used to reveal classicality versus true need for quantum computers.
In the past few years, the developments of practically-computable measures to characterize and quantify these aspects, 
{\it e.g.}~\cite{Amico:2007ag, Leone:2021rzd, Cao:2024nrx, Sierant:2025fax, Coffman:2025, Crew:2025fjg, Sarkis:2025cic}
has led to numerous studies of quantum complexity beyond entanglement in systems relevant to condensed matter, {\it e.g.}~\cite{Lamma:2026, Tarabunga:2024, Zavatti:2026, Frau:2024, Tirrito:2024, Odavic:2023, Russomanno:2025, Lakkaraju:2026}, 
quantum chemistry~\cite{Gu:2023ylw,Sarkis:2026, Seibert:2026}, 
lattice gauge theories~\cite{Ebner:2025pdm,Ebner:2026kwq,Minar:2026hqx, Falcao:2025, Tarabunga:2023, Froland:2026} ,
nuclear physics~\cite{Brökemeier:2025, Robin:2025, Robin:2025wip, Robin:2025dgw, Chernyshev:2025,Kadam:2026srm}
and high-energy phenomenology~\cite{White:2024bjp,Liu:2025qfl,Busoni:2025dns,Gargalionis:2025iqs,Gargalionis:2026onv,Li:2026zum,Liu:2026gju,Beane:2026gkh} 
(for a recent review, see~\cite{Robin:2026lqp}), 
as well as gravity~\cite{Cao:2024nrx, Grieninger:2026}. 
The acquired knowledge is starting to impact the developments of new classical techniques.
In the context of ground state finding, these include, for example,
stabilizer tensor networks~\cite{PhysRevLett.125.241602,PhysRevB.105.104306,Masot-Llima:2024doz,Nakhl:2024gfr}, Clifford-augmented matrix product states~\cite{PRXQuantum.6.010345,PhysRevLett.133.150604,Qian:2024vea,PhysRevLett.134.150403,PhysRevLett.134.150404,Huang:2024ron},
techniques based on Clifford Hamiltonian transformation~\cite{Mishmash:2023eow,Anand:2023ofu,Gu:2024ure}, non-stabilizerness injection~\cite{10.1145/3567955.3567958,Bhattacharyya:2023xad,Gu:2023ylw}, as well as non-contextual methods~\cite{Kirby2021contextualsubspace,Weaving:2022hbz,Seibert:2026} applied to quantum chemistry.
Further, the use of single stabilizer states that can be used as starting point for quantum refinement have been explored in Refs.~\cite{Sun:2024lts,Robin:2025wip,Mao:2026bjq}, and, very recently, superpositions of stabilizer states have been studied in models relevant to condensed matter physics in Ref.~\cite{Filip:2026}.

In order to further understand the relationships between physical aspects of many-body systems and features of quantum complexity,
studying models of increasing complexity provides a valuable framework for isolating specific phenomena and understanding their origins in fundamental aspects of quantum information, before progressing to fully realistic QMB systems.
The Lipkin–Meshkov–Glick (LMG) model~\cite{Meshkov:1965}, for example, can be formulated as a fermionic or spin model with infinite-range all-to-all interactions, and has been extensively studied because it exhibits spontaneous parity-symmetry breaking, analogous to nuclear deformation. It also provides an effective description of systems such as 
two-mode Bose–Einstein condensates~\cite{PhysRevA.55.4318,PhysRevA.67.013607,Fuentes-Schuller:2006xby}, 
trapped ion systems~\cite{Molmer:1998mq,PhysRevA.66.042101,PhysRevLett.90.133601,Lee:2014ftq}, 
and cavity QED setups~\cite{PhysRevA.77.043810}, 
and can generate spin-squeezed states relevant for quantum sensing and metrology~\cite{PhysRevA.68.012101,PhysRevA.80.012318,PhysRevA.90.022111,Lee:2014ftq,Calixto_2021}.

Entanglement aspects of this model have been widely studied, see {\it e.g.}~\cite{Wang:2003,Tullio:2019,Louren:2020,Calixto:2021, Faba:2021,Faba:2021b, Faba:2022, Hengstenberg:2023}, and more recently, non-stabilizerness~\cite{Passarelli:2024, Robin:2025wip}. 
It was found that the symmetry-broken regime is characterized by high, genuine multipartite entanglement, and low non-stabilizerness, 
suggesting that the associated collective particle-hole correlations and spin order are "simple", classically tractable phenomena.
These findings were then leveraged to efficiently approximate the LMG ground state with stabilizer states, yielding an improved initial state for quantum imaginary time evolution~\cite{Robin:2025wip}.

Besides particle-hole excitations, pairing is another fundamental type of correlation in fermionic systems that is worth examining in the context of quantum information~\cite{Kraus:2008zcr}. In nuclei, deformation and superfluid pairing constitute the two primary forms of collectivity, whose interplay and competition underlie many observed nuclear phenomena~\cite{Ring:1980}. 
Analogous competitions between pairing and symmetry-breaking collective orders also arise in other fermionic systems, including superconductors, ultracold atomic gases, and mesoscopic systems such as metallic clusters.
To study this interplay, the Agassi model~\cite{Agassi:1968} is an extension of the LMG Hamiltonian which includes pairing interactions and presents a richer phase diagram with both deformed and superfluid phases~\cite{Davis:1986, Garcia-Ramos:2018}.
Studies of quantum correlations and complexity in the Agassi model remain relatively scarce.
Ref.~\cite{Faba:2021dki} studied orbital quantum discord, while quantum simulations of this model have been implemented using both mappings to qubits~\cite{Saiz:2022rof,Perez-Fernandez:2021htx} and 5-dimensional qudits (qu5its)~\cite{Illa:2023scc}. 

In the present manuscript, we perform an extensive study of complexity in the Agassi Hamiltonian ground state, including measures of bi- and multi-partite entanglement, via entanglement entropy~\cite{Nielsen:2012}, $n$-tangles~\cite{Wootters:1998,Hill:1997,Coffmann:2000,Wong:2001} and quantum Fisher information (QFI)~\cite{RevModPhys.90.035005}, full-state and non-local non-stabilizerness via stabilizer R\'enyi entropies~\cite{Leone:2021rzd,Cao:2024nrx}, as well as fermionic non-Gaussianity via fermionic anti-flatness (FAF)~\cite{Sierant:2025fax}. We exploit these findings to extend the technique from Ref. ~\cite{Robin:2025wip} to the Agassi model including the pairing interaction to approximate the ground state by superpositions of stabilizer states in combination with a basis rotation.

The paper is organized as follows: 
Sec. \ref{Sec:Agassi} introduces the Agassi model and explains the mapping to qubits. 
In Sec. \ref{Sec:Complexity}, we study the complexity of the exact ground state of the Agassi model in terms of entanglement (Sec. \ref{sec:entang}), non stabilizerness (Sec. \ref{sec:magic} and \ref{sec:NLmagic}), and non-gaussianity (Sec. \ref{sec:non-gaussianity}). 
Sec. \ref{sec:stabgs} discusses the partitioning of the Hamiltonian into stabilizer pieces (Sec.~\ref{sec:stab_Hami}), the preparation of stabilizer states from graph states (Sec. \ref{sec:stabsfromgraph}), a ground state approximation using a superposition of stabilizer states (Sec. \ref{sec:stabexpansion}) and mode-pair basis rotation to improve this approximation (Sec. \ref{sec:br-stabexp}). 
Finally, Sec. \ref{sec:conclu} summarizes this work and discusses possible future applications.

%%%%%%%%%%%%%%%%%%%%%%%%%%%%%%%%
\section{The Agassi Hamiltonian and Mapping to Qubits}
\begin{figure}[h]
    \includegraphics[width=\columnwidth]{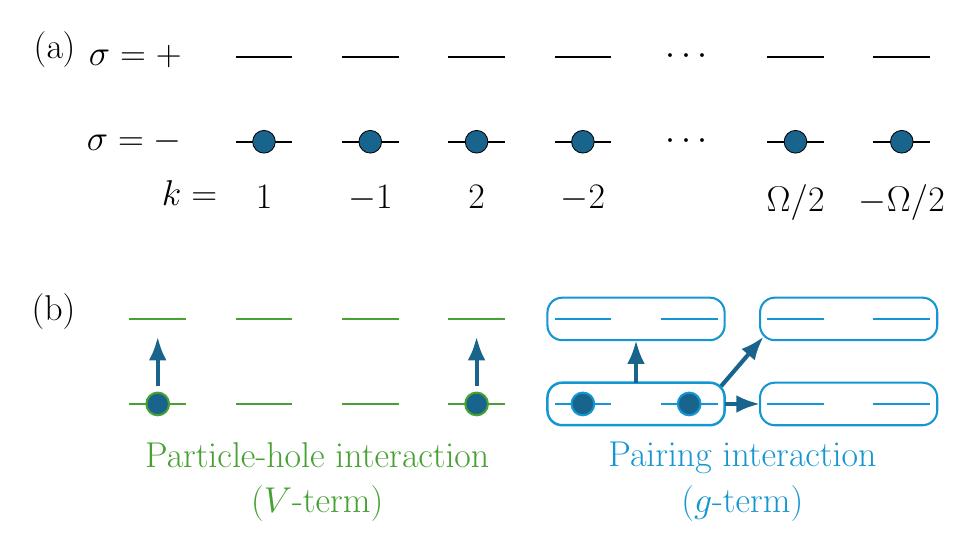}
    \caption{Illustration of the Agassi model in the half-filling case: $N=\Omega$ fermions are distributed on two levels $\sigma=\pm$, both $N$-fold degenerate. 
    Upper panel (a): The lowest-energy non-interacting (0p-0h) configuration of a system of $N$ particles in the Agassi model. 
    Lower panel (b): Possible interactions corresponding to the particle-hole and pairing terms in the Hamiltonian in Eq.~\eqref{eq:agassiDEF}, with strengths $V$ and $g$, respectively.}
    \label{fig:Agassi_model}
\end{figure}

\noindent
\label{Sec:Agassi}
The Agassi model~\cite{Agassi:1968} describes a system of $N$ interacting fermions distributed on two levels $\sigma=\pm$ with degeneracy $\Omega$, that are separated by energy $\varepsilon$, see Fig.~\ref{fig:Agassi_model}. 
In this model the fermions interact through both a particle-hole and pairing interaction.
Precisely, the Hamiltonian $\hat{H}$ governing the system is written as
\begin{align}
H & = 
 \varepsilon\ \hat S_z
\ - \ \frac{V}{2} \left( \hat{S}_+^2 + \hat{S}_-^2 \right) 
\ -\ 
g \hat B^\dagger \hat B %\\
%&=& \varepsilon \sum
\ ,
\label{eq:agassiDEF}
\end{align}
using
\begin{align}
\hat S_\alpha & =  \sum_{k=1}^{\Omega/2} \hat T_{k,\alpha}
\ \ ,\ \ 
\hat B \ =\ \sum_{k=1}^{\Omega/2} (\hat b_{k,+} + \hat b_{k,-}) \; ,
\end{align}
with
\begin{align}
%%%
\hat T_{k,+} & =   
\hat c_{k,+}^\dagger \hat c_{k,-} + 
\hat c_{-k,+}^\dagger \hat c_{-k,-}
\ ,\ \ 
%%%
\hat T_{k,-} \ =\ \hat T_{k,+}^\dagger
\ ,
\nonumber\\
%%%
\hat T_{k,z} & =  
 \frac{1}{2} \sum_{\sigma=\pm} \sigma \left( 
\hat c_{k,\sigma}^\dagger \hat c_{k,\sigma} 
+ \hat c_{-k,\sigma}^\dagger \hat c_{-k,\sigma} 
\right)
\nonumber\\
%%%
\hat b_{k, +} & =   
\hat c_{-k,+} \hat c_{k,+}
 \ ,\ \
 %%%
\hat b_{k, -} \ =\  \hat c_{-k,-} \hat c_{k,-} \ ,
\nonumber\\
%%%
\hat b_{k,z} & =   
{\frac{1}{\sqrt{2}} (
\hat c_{-k,+} \hat c_{k,-}
+ \hat c_{-k,-} \hat c_{k,+}
 ) }\ ,
\nonumber\\
%%%
\hat b^\dagger_{k, +} & =   
\hat c^\dagger_{k,+} \hat c^\dagger_{-k,+}
 \ ,\ \
%%%
\hat b^\dagger_{k, -} \ =\  \hat c^\dagger_{k,-} \hat c^\dagger_{-k,-} \ ,
\nonumber\\
%%%
\hat b^\dagger_{k,z} & =  
{\frac{1}{\sqrt{2}} (
\hat c^\dagger_{k,-} \hat c^\dagger_{-k,+}
+ \hat c^\dagger_{k,+} \hat c^\dagger_{-k,-}
 ) }\ ,
\end{align}
where $c^\dagger_{k,\sigma}$ ($c_{k,\sigma}$) creates (annihilates) a fermion in state $(k,\sigma)$. 
Adding the number operator
\begin{align}
\hat N_{k} & =  \sum_\sigma  \left(
\hat c_{k,\sigma}^\dagger \hat c_{k,\sigma} +
\hat c_{-k,\sigma}^\dagger \hat c_{-k,\sigma} \right) 
\; ,
\end{align}
completes a set of $10$ generators of SO($5$). 

In the following, we will refer to as "mode" $k$ a doublet of states $(k,\pm)$, while a pair of fermions refers to two fermions in the paired states $(k,\sigma)$ and $(-k,\sigma)$ of the same level $\sigma$. 
Finally a mode-pair is a set of four single particle states $(k,\pm)$, $(-k,\pm)$.
The particle-hole (monopole) interaction ($V$) in Eq.~\eqref{eq:agassiDEF} scatters particles from one level to the other within the same mode $k$, while the pairing interaction ($g$) can move pairs of fermions between modes within the same level or to the other level, see lower panel of Fig. \ref{fig:Agassi_model}. 

Depending on the dominant parameter ($\varepsilon$, $V$ or $g$), the system transitions between three different phases that were studied through mean-field calculations in the half-filling case, for example, in Ref. \cite{Davis:1986, Garcia-Ramos:2018}. For small interaction strengths, $V$ and $g$, compared to $\varepsilon$, the system is in the normal phase which can be described by the spherical Hartree-Fock solution. 
When $V$ dominates, the particle-hole interaction is the main mechanism for introducing correlations and the system enters a parity-broken deformed phase, with a mixed-parity mean-field solution.
Finally, if $g$ is larger than the other parameters, the system is in the superfluid Bardeen–Cooper–Schrieffer (BCS) phase, with degenerate ground states of different particle numbers.
This rich phase structure makes the Agassi model interesting for different topics in physics and quantum chemistry.

When no pairing interaction is present ($g=0$), the Agassi model reduces to the Lipkin-Meshkov-Glick (LMG) model~\cite{Lipkin:1965}. 
While the LMG model can be interpreted as a spin system (because the monopole interaction preserve the single occupancy within each mode $k$), this is not the case for the Agassi model, as the pairing interaction allows modes to be completely filled or empty.
\\

\begin{figure}[h]
    \centering
    \includegraphics[width=0.7\linewidth]{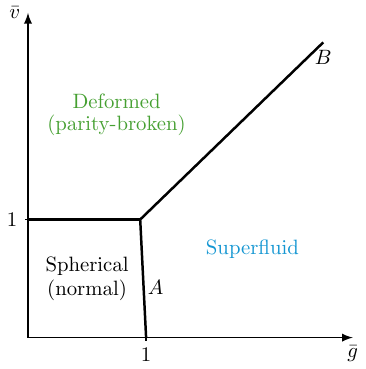}
    \caption{The phase diagram of the Agassi model for $N=\Omega$, where the phase transition $A$ is given by $\vbar=(\Omega-1)(1-\gbar)$ and $B$ corresponds to $\vbar=\frac{\Omega-1}{\Omega-2}\gbar$. Adapted from Ref.~\cite{Davis:1986}. } 
    \label{fig:phasediagram}
\end{figure}

In the following we will work in the half-filling case, {\it i.e.} with a number of fermions $N = \Omega$. For convenience, we will use the dimensionless Hamiltonian $\widetilde{H} = \hat{H}/\varepsilon$ and re-scaled parameters
\begin{align}
    \vbar &= \frac{V (\Omega-1)}{\varepsilon} \ ,\nonumber\\
     \gbar &= \frac{g (\Omega-1)}{\varepsilon} \ .
\end{align}
The relevant Hamiltonian thus reads
\begin{align}
 \widetilde{H} 
& = \hat S_z
-\frac{\vbar}{2 (\Omega -1)} \left( \hat{S}_+^2 + \hat{S}_-^2 \right) 
- \frac{\gbar}{ (\Omega -1)} \hat B^\dagger \hat B \nonumber \\
& \equiv \widetilde{H}_0 + \widetilde{H}_{v} +  \widetilde{H}_{g} \; ,
\label{eq:rescaled_Hami}
\end{align}
As a reminder we show the corresponding phase diagram obtained from mean-field calculations~\cite{Davis:1986} in Fig.~\ref{fig:phasediagram}.

\begin{figure}[!h]
\centering{\includegraphics[width=0.45\columnwidth] {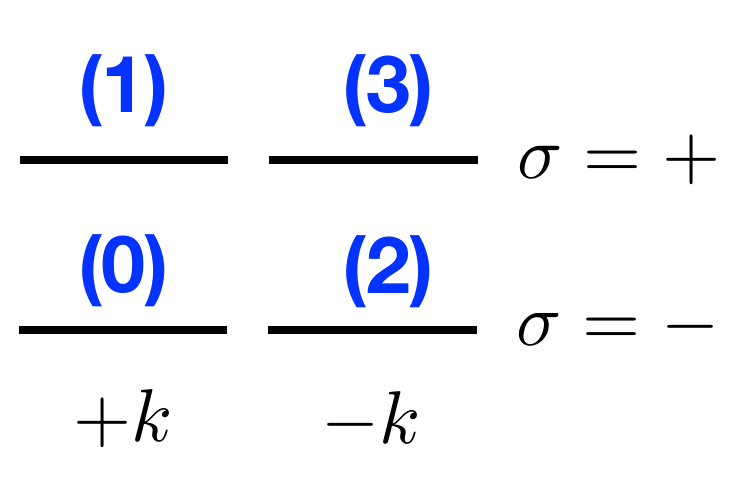}} 
\caption{pa-JW mapping of a mode-pair to four qubits $m=0,1,2,3$.}
\label{fig:PAJW_mapping}
\end{figure}

In this study we choose a simple Jordan-Wigner mapping of single-fermion states onto qubits~\footnote{Ref.~\cite{Illa:2023scc} showed that the Agassi model can naturally be mapped onto 5-dimensional qudits, or qu5its, due to the underlying $so(5)$ algebra. In this work we 
focus on a qubit mapping, 
which will provide a point of comparison for studies of quantum complexity with qu5it mappings, which we leave to a future work.
}.
Specifically we adopt the {\it physics-aware} Jordan-Wigner (pa-JW) mapping of Ref.~\cite{Illa:2023scc}, 
with convention
\begin{align}
    c_m &= \left( \prod_{l<m} -Z_l \right) \sigma_m^- \; , \nonumber \\
    c_m^\dagger &= \left( \prod_{l<m} -Z_l \right) \sigma_m^+ \; ,
    \label{eq:mapping}
\end{align}
where $\sigma_m^\pm = (X_m \pm i Y_m)/2$, with $X,Y,Z$ denoting the usual Pauli operators.
In the pa-JW mapping, the qubits are ordered as shown in Fig.~\ref{fig:PAJW_mapping} to exploit the bosonic nature of the mode pairs. 
Precisely, adjacent qubits are grouped within mode pairs, and thus, as the number of particles in a given mode pair remains even, there are no fermionic phases from operators moving from one mode pair to another, which minimizes the number of $Z$ operators.
According to this mapping, the Hamiltonian in Eq. \eqref{eq:agassiDEF} can be rewritten in terms of Pauli strings acting on qubits. The full expression is given in appendix~\ref{app:Hamiltonian}.
Since each single particle state $(k,\sigma)$ is mapped onto one qubit, $N_q = 2N$ qubits are required to describe a system with $N=\Omega$ particles. 

%%%%%%%%%%%%%%%%%%%%%%%%%%%%%%%%
%%%%%%%%%%%%%%%%%%%%%%%%%%%%%%%%
\section{Complexity in the Exact Agassi Ground State}
%%%%%%%%%%%%%%%%%%%%%%%%%%%%%%%%
%%%%%%%%%%%%%%%%%%%%%%%%%%%%%%%%
\noindent
\label{Sec:Complexity}
The complexity in a many-body quantum state can be characterized through various aspects.
Concepts from information theory, such as entanglement, non-stabilizerness, and non-Gaussianity provide a precise way to relate the traditional concepts of many-body correlations to computational resources.
While entanglement is known to be necessary to make a quantum state complex, in the sense that it cannot be represented by a polynomial number of classical configurations, it is also not sufficient. 
Non-stabilizerness or non-gaussianity, then constitute the "magic" ingredient that makes a quantum computer truly required to simulate the state~\cite{Gottesman:1998, Bartlett:2002}.

If one of these three resources is small, the state can be efficiently described classically, if the entanglement is limited, the state can be described using tensor networks~\cite{White:1992, Schollwoeck:2011, Shi:2006, Verstraete:2004, Vidal:2008}, the stabilizer code allows to simulate states with small non-stabilizerness~\cite{Aaronson:2004, Bravyi:2016} and states with vanishing non-gaussianity can be characterized at low cost by their covariance matrix~\cite{Hebenstreit:2019,Bravyi:2005,Dias:2024}. In the following we will discuss different measures for these resources to quantify the complexity of the ground state of the Agassi model.

%%%%%%%%%%%%%%%%%%%%%%%%%%%%%%%%
%%%%%%%%%%%%%%%%%%%%%%%%%%%%%%%%
\subsection{Entanglement}
\label{sec:entang}
%%%%%%%%%%%%%%%%%%%%%%%%%%%%%%%%
%%%%%%%%%%%%%%%%%%%%%%%%%%%%%%%%
\noindent

We start by investigating various aspects of entanglement of the Agassi model, from simple bi-partite entanglement entropies to multipartite measures.
While the entanglement properties of the LMG ground state have been extensively studied, see e.g.~\cite{Faba:2021,Faba:2021b, Faba:2022, Wang:2003,Tullio:2019,Louren:2020,Calixto:2021, Hengstenberg:2023}, to the best of our knowledge, extensions to the Agassi Hamiltonian have so far been limited to calculations of the quantum discord~\cite{Faba:2021dki}  \\

\begin{figure}[h!]
     \centering
    \includegraphics[width=\linewidth]{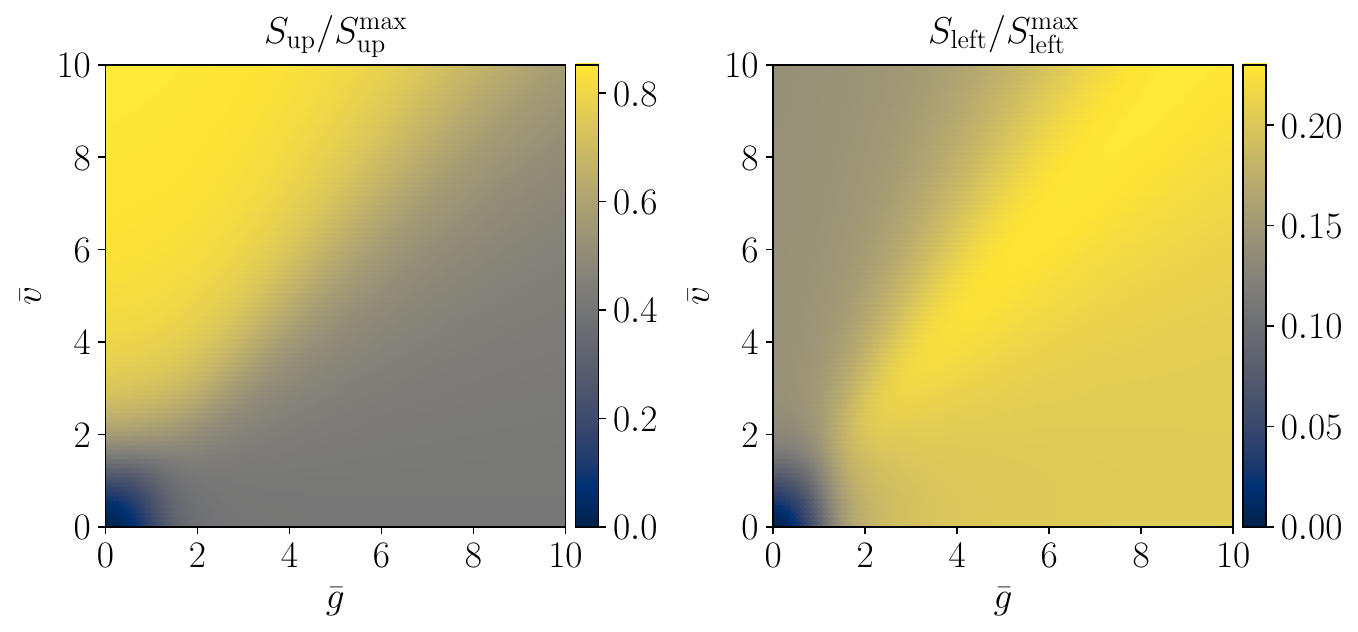}
    \caption{Ratios of von Neumann entanglement entropies over the corresponding maximal value $S_A/S_A^{\rm max}$ for the left/right bi-partitioning (left panel), and up/down bi-partitioning (right panel), as a function of interaction parameters $\vbar$ and $\gbar$. Note the different scales on the plot.}
    \label{fig:entang_entropies}
\end{figure}

In computing the von Neumann entanglement entropies
\begin{align}
    S_A = - {\rm Tr} \left( \rho_A \log_2 \rho_A\right) \; ,
\end{align}
where $\rho_A = {\rm Tr}_B (\rho_{AB})$,
we focus on two types of bi-partitionings into $A/B$ sub-subsystems: a) a left/right partitioning in which $L=\lfloor \Omega/4 \rfloor$ mode-pairs are traced out, and b) an up/down partitioning in which one of the two levels is traced out.
Results for a system with $\Omega=8$ ($N_q=16$) are displayed in Fig.~\ref{fig:entang_entropies}.
Overall we found $S_{\rm up}$ to be largely driven by the particle-hole interaction strength $\vbar$, with $S_{\rm up} \gtrsim 0.8 \, S_{\rm up}^{\rm max}$ in the LMG limit ($\gbar = 0$) above the phase transition, while $S_{\rm left}$ is mostly caused by the pairing interaction but remains well below the maximal value.
As described in more details in appendix~\ref{app:entanglement}, the entanglement between up and down levels is extensive in the LMG limit, {\it i.e.} $S_{\rm up} \propto \Omega$, while the scaling is slightly suppressed when pairing is introduced.  On the other hand $S_{\rm left}$ is constant in the LMG limit and exhibits sub-linear scaling in the pairing limit.
\\

Beyond bi-partitions, measures of multi-partite entanglement provide a more detailed characterization of the entanglement structure of the system. However, quantifying genuine $n$-partite entanglement remains a challenging problem, and thus, unlike in the bi-partite case, there is no universally accepted measure that fully captures all aspects of multipartite correlations, and different approaches often emphasize different facets of entanglement.
In this work we explore the $n$-tangles \cite{Hill:1997, Wootters:1998, Coffmann:2000, Wong:2001}, which describe how $n$ qubits are entangled with each other in the larger $N_q$-qubit system that maps the Agassi ground state. 
For even values of $n$, the $n$-tangles are defined as
\begin{align}
    \tau^{(n)}_{i_1,...,i_n} = |\langle \Psi | Y_{i_1} \otimes ... \otimes Y_{i_n}   | \Psi^* \rangle|^2 \; ,
\label{eq:ntangles}
\end{align}
where $Y_{i_k}$ is the Pauli operator $Y \equiv \hat{\sigma}_y$ acting on qubit $i_k$. 
These $n$-tangles are related to $2n$-fermion entanglement, through the JW mapping in Eq.~\eqref{eq:mapping}~\footnote{We note, however, that the definition in Eq.~\eqref{eq:ntangles} has been designed for qubit systems and does not account for fermionic anti-symmetry. This makes the values dependent on the chosen qubit ordering. }.

\begin{figure}[h]
    \centering
    \subfloat[Type-1: $\tau^{(4)}_{0123}$]{\includegraphics[width=0.49\linewidth]{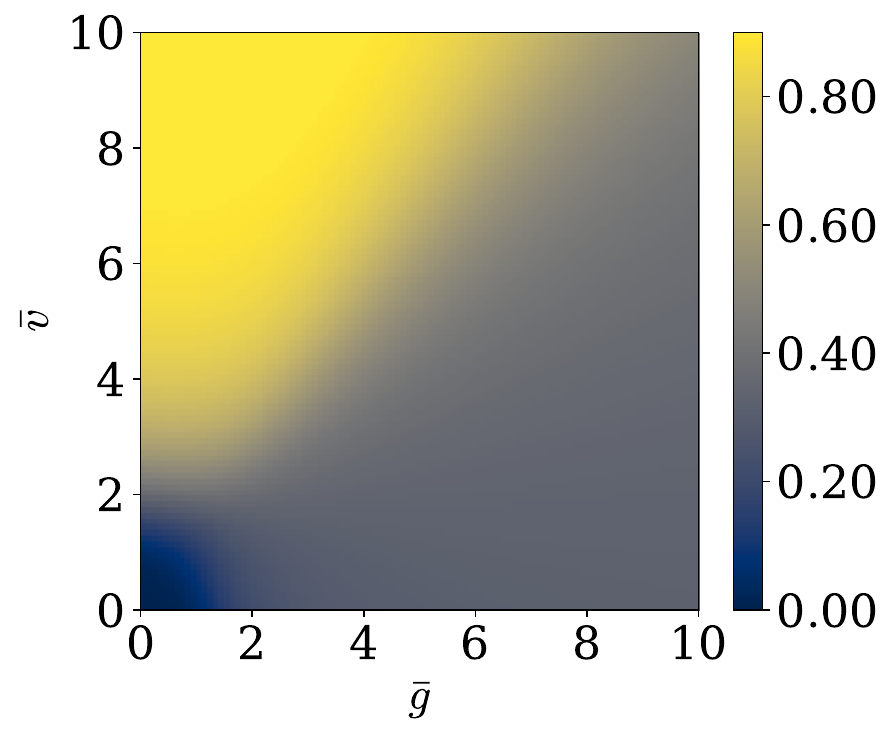}}
    \subfloat[Type-2: $\tau^{(4)}_{0145}$]{\includegraphics[width=0.49\linewidth]{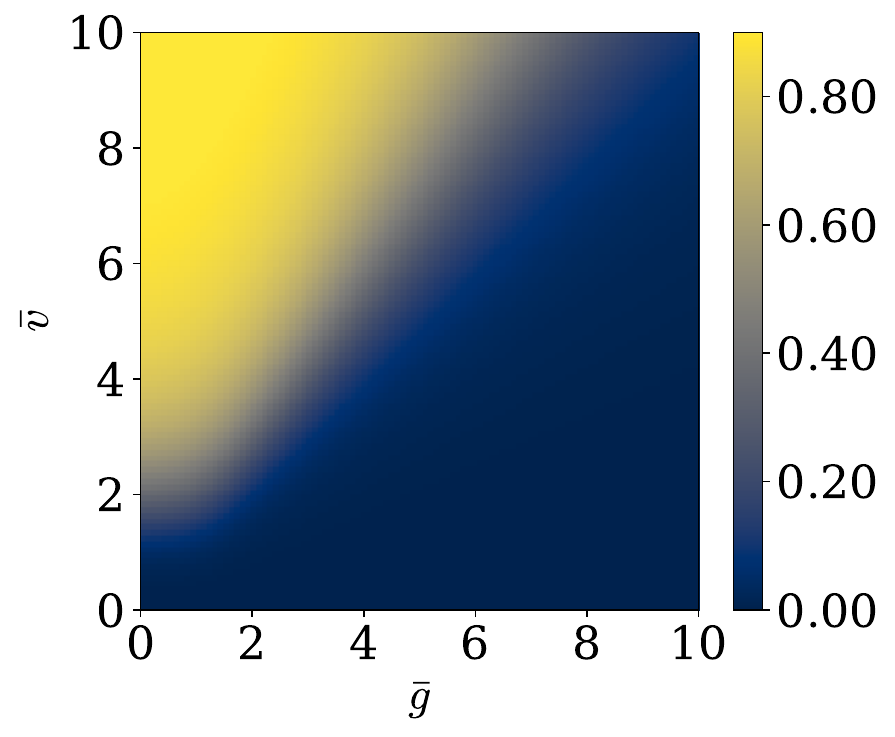}}
    \hfill
    \subfloat[Type-3: $\tau^{(4)}_{0246}$]{\includegraphics[width=0.49\linewidth]{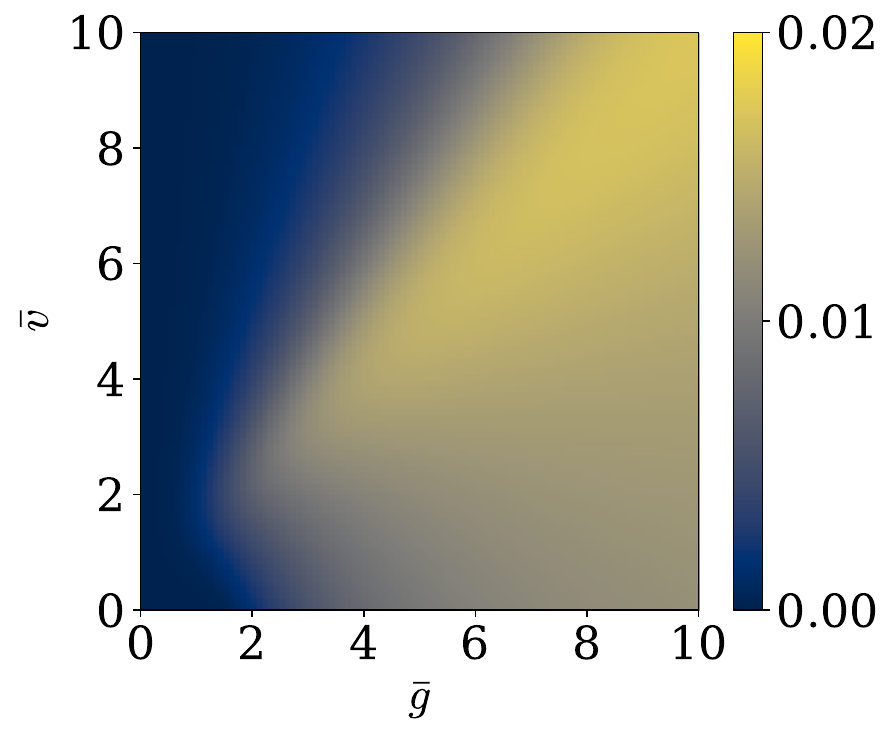}}
    \subfloat[Type-4: $\tau^{(4)}_{0257}$]{\includegraphics[width=0.49\linewidth]{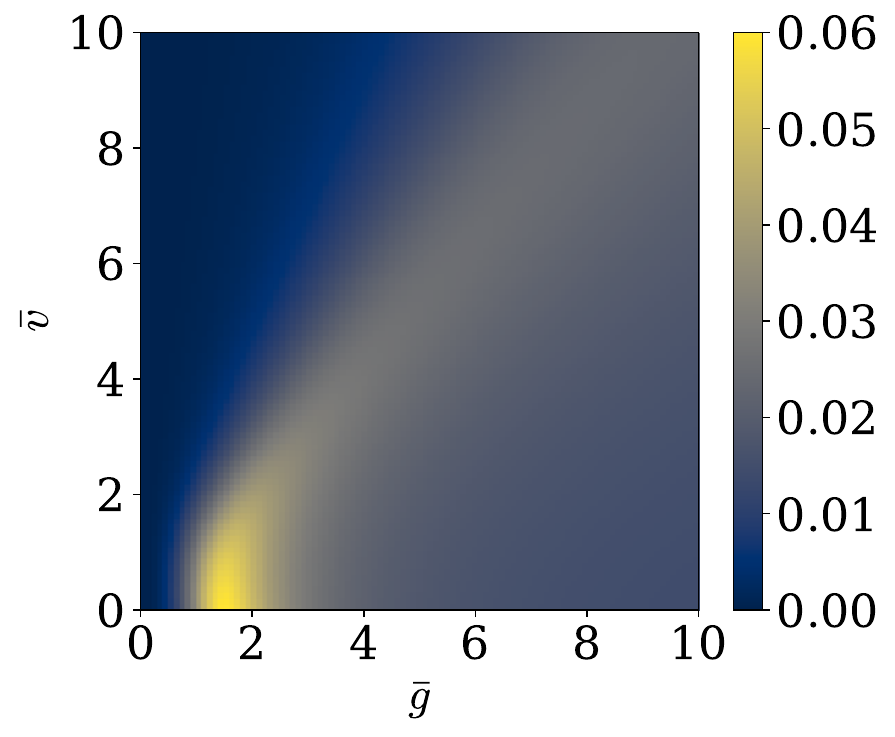}}
    \hfill
    \subfloat[Type-5: $\tau^{(4)}_{0347}$]{\includegraphics[width=0.49\linewidth]{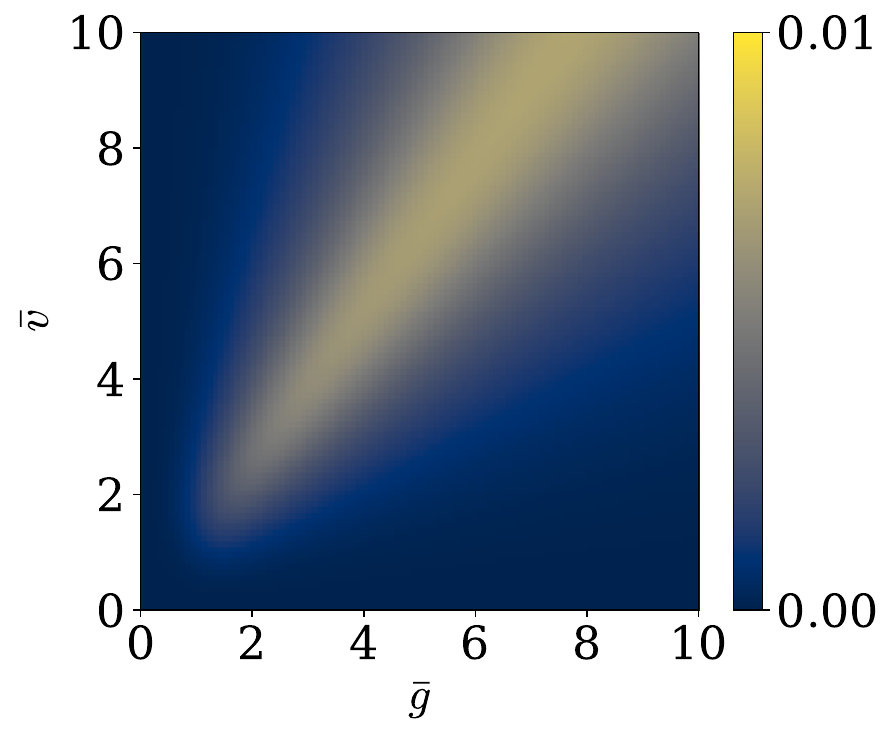}}
\caption{All types of $4$-tangles for a system $\Omega=8$. Note the different scales.}
\label{fig:4tangles_Omega8}
\end{figure}

In the present Agassi model, it is found that 2- and 3-tangles
cancel due to symmetries,
and thus the minimal tangles occur among 4 qubits (for example, 2 modes or 2 pairs). 
Due to the permutation symmetry in terms of mode pairs, there are essentially five types of 4-tangles that are relevant for arbitrary $\Omega$, for example:
\begin{itemize}
    \item Type 1: 4-tangles involving 4 sites of the same mode-pair, {\it e.g.} $\tau^{(4)}_{0123}$.
    \item Type 2: 4-tangles involving two modes of different mode-pairs: {\it e.g.} $\tau^{(4)}_{0145}$.
    \item Type 3: 4-tangles involving two pairs of states on the same level, {\it e.g.} $\tau^{(4)}_{0246}$ or $\tau^{(4)}_{1357}$.
    \item Type 4: 4-tangles involving two pairs of states on different levels, {\it e.g.} $\tau^{(4)}_{0257}$ or $\tau^{(4)}_{1346}$.
    \item Type 5: 4-tangles involving two states on different levels and different modes per mode-pair, {it e.g.} 
    $\tau^{(4)}_{0347}$.
    
\end{itemize}
Fig. \ref{fig:4tangles_Omega8} displays these different types of $4$-tangles for a system with $\Omega=8$.
It is seen that $4$-qubit entanglement is largest in the deformed phase, with main contributions of type-1 and type-2 4-tangles. While type-1 tangles are also present in the superfluid phase (to a much smaller extent), the $4$-tangles of type-2 are not, since the corresponding single-particle orbitals are mostly connected through the particle-hole interaction. 
The opposite holds for $4$-tangles of type-3 and type-4, which are basically absent from the deformed phase. These types of tangles are overall characterized by very small magnitudes, and mostly arise from the pairing interaction. As seen in Fig.~\ref{fig:4tangles_Omega8}, type-3 and type-4 tangles are also more pronounced in the regions close to the transitions to deformed and normal phases, respectively.
The $4$-tangles of type-5 are essentially vanishing in the deformed and superfluid phase since one of the two interactions cannot connect the corresponding orbitals alone.

Beyond 4-tangles, one can in principle calculate $n$-tangles with $n>4$ to probe higher-order multi-partite entanglement. For example, we find the maximum-order $2\Omega$-tangles to be uniformly maximal across both the deformed and superfluid regimes, and in the region of transition between the two.
We remind, however that these $n$-angles generally include contributions from lower-order $m$-tangles ($m<n$) and therefore do not provide a clean measure of genuine $n$-partite entanglement.
\\

To complement these measures, the Quantum Fisher Information (QFI)~\cite{RevModPhys.90.035005} can be used as a witness for genuine $n$-partite entanglement~\cite{Hyllus:2012ufd,Toth:2012lpv}.
Originating from quantum metrology, the QFI provides a measure of the sensitivity of a many-body quantum state to collective transformations generated by a given operator $\hat{O}$.
For a pure state $\ket{\Psi}$, the QFI $F_Q$ reduces to four times the variance of that operator~\cite{RevModPhys.90.035005}
\begin{align}
    F_Q = 4 \left( \langle \Psi | {\hat O}^2 | \Psi \rangle  - \langle \Psi | {\hat O}  | \Psi \rangle^2 \right) \; .
\end{align}
It is known that for a $m$-producible state (a state with no genuine $(m+1)$-partite entanglement~\footnote{A state is $m$-producible if it can be written as a tensor product of states, each involving at most $m$ physical subsystems.}), $F_Q$ is bounded by~\cite{Hyllus:2012ufd,Toth:2012lpv}
\begin{align}
    F_Q \leq F_{\rm lim} = s m^2 + r^2 \; ,
\
\end{align}
where $s = \lfloor \mathcal{N}_{\rm phys}/m\rfloor$, $r=\mathcal{N}_{\rm phys} - sm$, and  $\mathcal{N}_{\rm phys}$ denotes the number of physical degrees of freedom.
Thus if $F_Q$ exceeds that bound, {\it i.e.} if $F_Q > F_{\rm lim}$, then $\ket{\Psi}$ is at least $(m+1)$-partite genuinely entangled.

Generators $\hat O$ are typically chosen as those exhibiting the largest quantum fluctuations. In the deformed phase of the Agassi model, which is dominated by the LMG interaction $\hat{H}_v$ in Eq.~\eqref{eq:rescaled_Hami}, a natural choice for $\hat O$ is given by $\hat{J}_x = (\hat{J}_+ + \hat{J}_-)/2$, which is favored over $\hat{J}_y$ in the present anisotropic case. In the superfluid regime, on the other hand, the pairing interaction $\hat{H}_g$ dominates and thus
natural choices are $\hat{B}_{x} = (\hat{B}^\dagger + \hat{B})/2$ or $\hat{B}_{y} = (\hat{B}^\dagger - \hat{B})/(2i)$. 
In the following we calculate the QFI obtained from all of these generators to determine the {\it entanglement depth}~\cite{PhysRevLett.86.4431} $n=m+1$, which certifies genuine $n$-partite entanglement, across the Agassi phase diagram. 
Since these operators are local in term of modes or pairs, each consisting of two orbitals, the number of physical subsystems here is $\mathcal{N}_{\rm phys}  = \Omega$.

\begin{figure}[h!]
     \centering
    \includegraphics[width=0.8\linewidth]{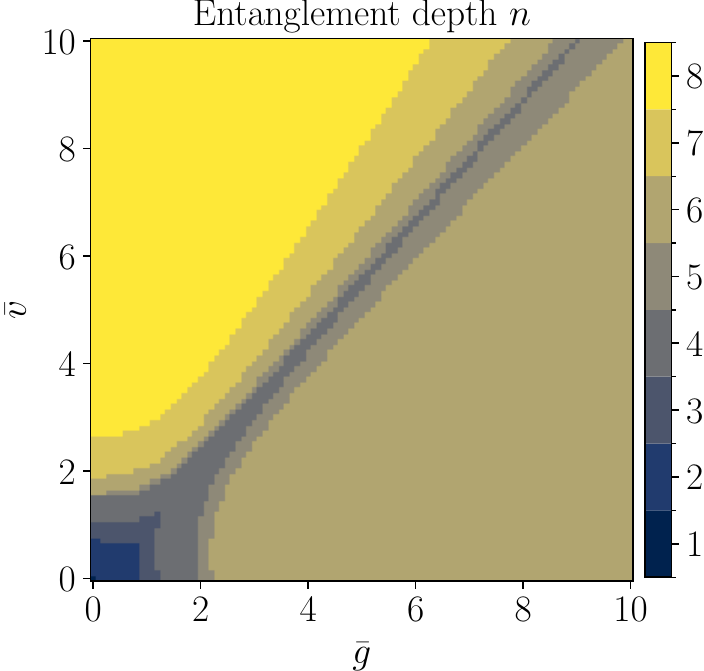}
    \caption{Entanglement depth $n$ certifying genuine $n$-partite entanglement, where each party refers to a mode or a pair (two orbitals), for a system with $\Omega=8$. }
    \label{fig:QFI_Ome8}
\end{figure}

Fig.~\ref{fig:QFI_Ome8} displays the entanglement depth $n=m+1$ for a system with $\Omega=8$.
In the normal phase, for $\gbar$ and $\vbar \lesssim 1-2$, the Agassi ground state presents little multi-partite entanglement, as already noted from the $n$-tangles.
The deformed phase, on the other hand, displays maximal entanglement depth $n=\Omega$, indicating that all $\Omega$ modes are genuinely collectively entangled with each other. This is seen for various values of $\Omega$~\footnote{Note that this observation was made within the LMG limit $\gbar=0$ in Refs.~\cite{Salvatori:2014gan,Robin:2025wip}, which showed close-to-maximal $2\Omega$-partite entanglement in the deformed phase of that model.}.
In the superfluid phase, while multi-partite entanglement is present, the entanglement depth is lower and indicates that only at least 6 pairs are genuinely entangled.   
Interestingly the entanglement depth is lower at the phase transitions. 
\\

To conclude, it appears from the above study that the particle-hole interaction ($\vbar$) yields large, extensive and genuinely collective many-body entanglement, while the pairing interaction ($\gbar$) leads to milder scaling of von Neumann entropies and less collective multi-partite entanglement.

%%%%%%%%%%%%%%%%%%%%%%%%%%%%%%%%
%%%%%%%%%%%%%%%%%%%%%%%%%%%%%%%%
\subsection{Non-Stabilizerness}
\label{sec:magic}
%%%%%%%%%%%%%%%%%%%%%%%%%%%%%%%%
%%%%%%%%%%%%%%%%%%%%%%%%%%%%%%%%
\noindent
Complementary to measures of entanglement, measures of non-stabilizerness aim at quantifying, in some way, the distance of a quantum state to the manifold of stabilizer states.
For an $N_q$-qubit system, such stabilizer states $\ket{\Psi_s}$ have the particular property that they are fully determined by their $2^{N_q}$-dimensional stabilizer group
\begin{align}
    \mathcal{S}(\ket{\Psi_s}) = \{ \hat{P} \in \mathcal{G}_{N_q}, \; \hat{P} \ket{\Psi_s} = \ket{\Psi_s} \} \; ,
\end{align}
which is an abelian subgroup of the generalized $N_q$-qubit Pauli group
\begin{align}
\mathcal{G}_{N_q} = \{\varphi \; \sigma^{(1)}\otimes\sigma^{(2)}\otimes...\otimes\sigma^{({N_q})}\} \; ,
\end{align}
where $\sigma^{(i)}\in\{I,X,Y,Z\}$ is a Pauli matrix acting on qubit $i$ and $\varphi\in\{\pm 1,\pm i\}$ is a phase.
This property means that a stabilizer state $\ket{\Psi_s}$ can be equivalently completely specified by the $N_q$ generators of $\mathcal{S}(\ket{\Psi_s})$, and thereby can be represented efficiently.

A practically-computable and commonly-used measure of non-stabilizerness in a $N_q$-qubit state $\ket{\Psi}$ is the stabilizer $\alpha$-R\'enyi entropy ($\alpha$-SRE), introduced in Ref.~\cite{Leone:2021rzd} as
\begin{eqnarray}
        {\cal M}_\alpha (\ket{\Psi}) & = &  -\log_2 \ d + \frac{1}{1-\alpha}\log_2 \ \sum_{\hat{P} \in \widetilde{\mathcal{G}}_{{N_q}}}  \left(\frac{\langle \Psi |\hat{P} | \Psi \rangle ^2}{d}\right)^\alpha \nonumber \; ,
\end{eqnarray}
where $d=2^{{N_q}}$ and $\widetilde{\mathcal{G}}_{{N_q}}$ is the subgroup of ${\mathcal{G}}_{{N_q}}$ with phase $\varphi=+1$.
Depending on the value of $\alpha$, the SRE probes different aspects of non-stabilizerness. 
Precisely, for $\alpha>1$, the SRE measures the distance to the closest stabilizer state, while for $\alpha<1$ the SRE relates to the stabilizer rank, {\it i.e.} the minimal number of stabilizer states needed in a superposition to represent $\ket{\Psi}$~\cite{Leone:2021rzd,Haug:2024ptu}. 
\\

\begin{figure}[ht]
    \centering
    \subfloat[Stabilizer $2$-R\'enyi Entropy $\mathcal{M}_2$]{\includegraphics[width=0.8\linewidth]{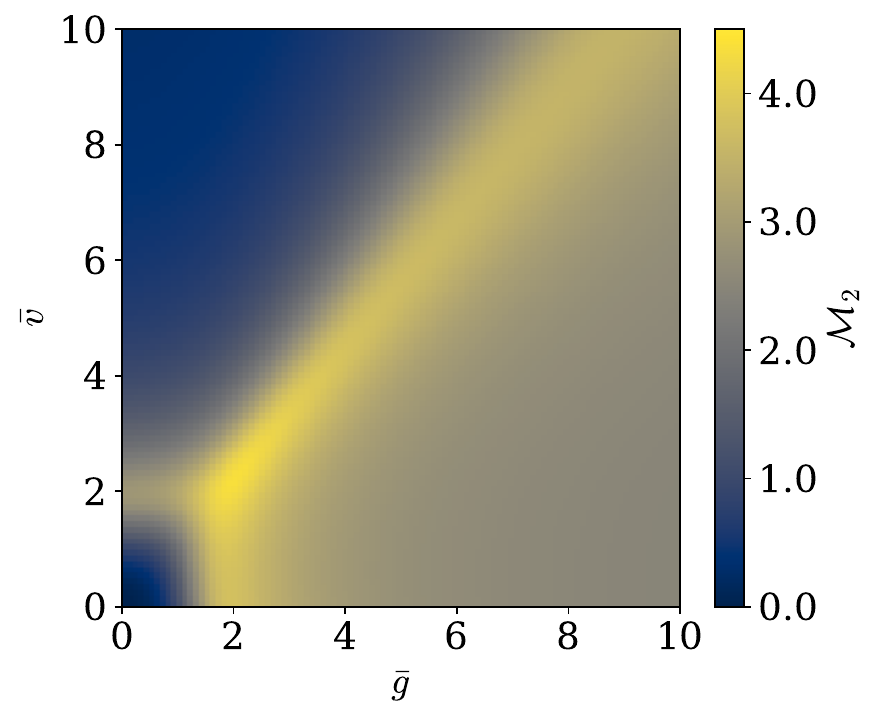}}
    \hfill
    \subfloat[Non-local Stabilizer $2$-R\'enyi Entropy $\mathcal{M}_2^{\rm{(NL)}}$ ]{\includegraphics[width=0.8\linewidth]{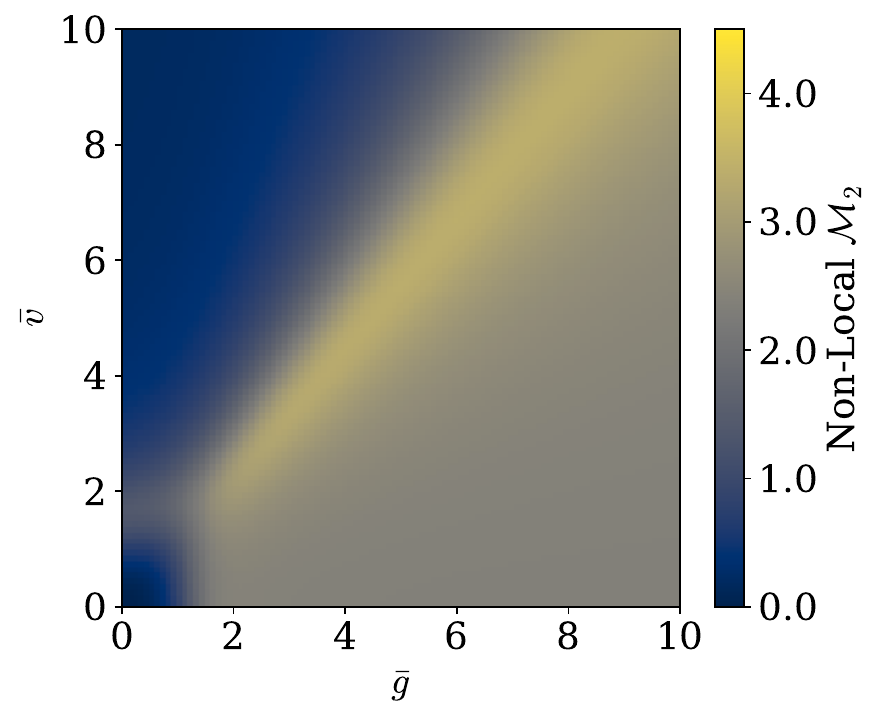}}
    \caption{Full-state and non-local 2-SRE, $\mathcal{M}_2$ and $\mathcal{M}_2^{\rm{(NL)}}$, for $\Omega=8$ as a function of $\vbar$ and $\gbar$. }
    \label{fig:M2_NLM2}
\end{figure}

\begin{figure}[ht]
    \centering
    \subfloat[Stabilizer $1/2$-R\'enyi Entropy $\mathcal{M}_{1/2}$]{\includegraphics[width=0.8\linewidth]{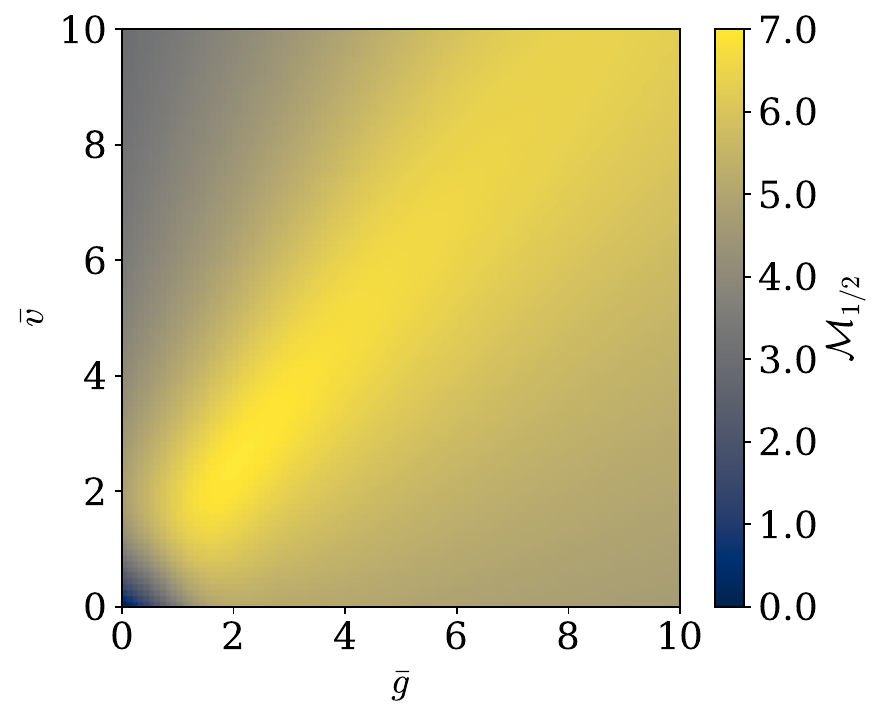}}
    \hfill
    \subfloat[Non-local Stabilizer $1/2$-R\'enyi Entropy $\mathcal{M}_{1/2}^{\rm{(NL)}}$ ]{\includegraphics[width=0.8\linewidth]{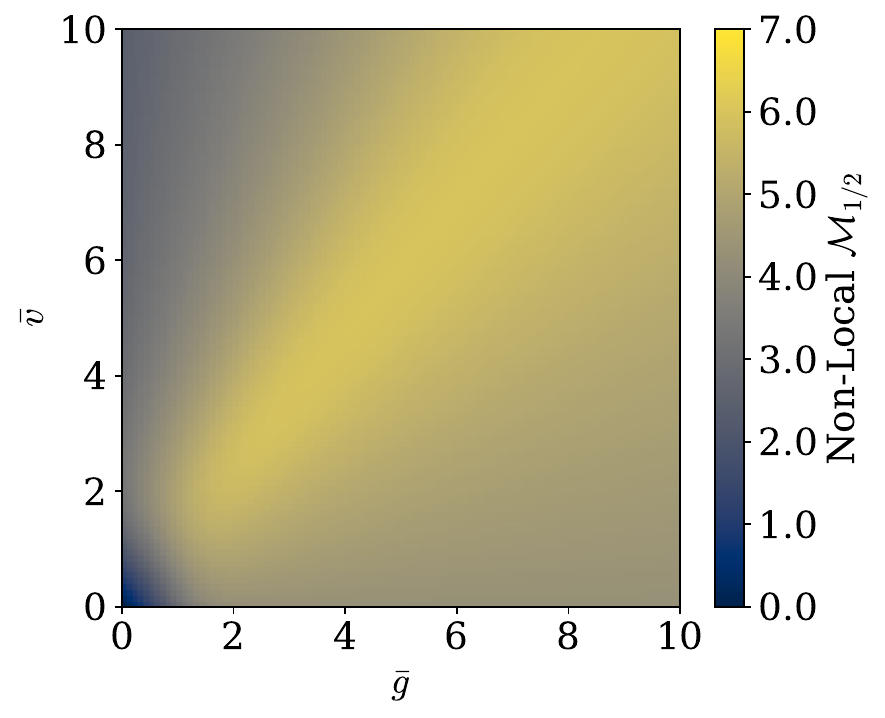}}
    \caption{Full-state and non-local $1/2$-SRE, $\mathcal{M}_{1/2}$ and $\mathcal{M}_{1/2}^{\rm{(NL)}}$, for $\Omega=8$ as a function of $\vbar$ and $\gbar$. .
    }
    \label{fig:Mhalf_NLMhalf}
\end{figure}

To illustrate both aspects, we choose to compute the $\alpha$-SRE of the Agassi ground state for $\alpha=2$ and $\alpha=1/2$. The results are shown for a system with $\Omega=8$ in 
%Fig.~\ref{fig:SREs_Ome8}.
Fig.~\ref{fig:M2_NLM2}(a) and~\ref{fig:Mhalf_NLMhalf}(a).
The system exhibits different characteristic patterns of 2-SRE $\mathcal{M}_2$ across the phase diagram, as
we observe three different regions corresponding to the three phases of the Agassi model. Specifically, The $2$-SRE is negligible in the normal and deformed phases, indicating that the ground state is very close to a single stabilizer state, while it remains sizable in the superfluid phase. 
At the phase transitions between these phases, the $2$-SRE displays "magic barriers" where $\mathcal{M}_2$ is extensive. More details on the scaling of $\M_2$ can be found in appendix \ref{app:NS}.
This contrasts, however with the behavior of multi-partite entanglement in Fig.~\ref{fig:QFI_Ome8}, for which the entanglement depth becomes smaller at the phase transitions.
\\
The $1/2$-SRE $\mathcal{M}_{1/2}$, in comparison, exhibits a more uniform pattern, as shown in %Fig.~\ref{fig:SREs_Ome8}.
Fig.~\ref{fig:Mhalf_NLMhalf}(a).
While $\mathcal{M}_{1/2}$ appears to grow approximately as $\sim \Omega^{0.46}$ in the deformed phase, it grows roughly as $\sim \Omega^{1.26}$ in the superfluid phase, see appendix \ref{app:NS}. 
\\
Together, the behaviors of $\M_2$ and $\M_{1/2}$ suggest that a superposition of a large number of stabilizer states will be needed to approximate the ground state in the superfluid regime. 
As we show in Sec.~\ref{sec:stabgs}, an exact representation of the ground state in fact requires a number of stabilizer states that grows exponentially with system size. Basis rotations nevertheless yield accurate starting points for quantum refinement using fewer stabilizer states, although their number still grows extensively with system size.
On the other hand, a single stabilizer state can provide a good approximation to the ground state in the deformed phase, while capturing it exactly would require a stabilizer rank that grows more slowly than in the superfluid phase.

%%%%%%%%%%%%%%%%%%%%%%%%%%%%%%%%
%%%%%%%%%%%%%%%%%%%%%%%%%%%%%%%%
\subsection{Non-Local Non-Stabilizerness}
\label{sec:NLmagic}
%%%%%%%%%%%%%%%%%%%%%%%%%%%%%%%%
%%%%%%%%%%%%%%%%%%%%%%%%%%%%%%%%
\noindent
Regimes in which both entanglement and non-stabilizerness are large indicate the presence of true quantum complexity, and the need for quantum computers to simulate the system at scale (within the same formulation and mapping).
The interplay between these two aspects can be further captured by the notion of non-local non-stabilizerness, which quantifies the non-stabilizerness that survives under changes of local bases.
Precisely, given a chosen $n$-partitioning of the system, this can be evaluated via non-local SREs, defined as~\cite{Cao:2024nrx}
\begin{align}
    \M_\alpha^{(n, \, \rm{NL})} (\ket{\Psi}) = \min_{\hat{U}_{A_1} \otimes ... \otimes \hat{U}_{A_n} } \M_\alpha (\hat{U}_{A_1} \otimes ... \otimes \hat{U}_{A_n} \ket{\Psi}) \; ,
\label{eq:NL_NS}
\end{align}
where the $U_{A_i}$ are unitaries acting on the chosen local degrees of freedom.

In general, investigating non-local non-stabilizerness for various partitionings may allow one to identify the relevant degrees-of-freedom and their optimal bases to further simulate the system at larger scales.
In the present Agassi model, the underlying $so(5)$ algebra provides an intuitive way to partition the system into mode-pairs~\cite{Illa:2023scc}.
Constraining the unitary operations on the mode-pairs to respect the symmetries of the model, such as particle number and bosonic nature of the mode pairs, significantly reduces the number of parameters to optimize. This is described in more details in appendix~\ref{app:NL_magic}.

Figs.~\ref{fig:M2_NLM2}(b) and~\ref{fig:Mhalf_NLMhalf}(b)
show the non-local 2-SRE and $1/2$-SRE for a system with $\Omega =8$, respectively.
It is seen that the 2-SRE and $1/2$-SRE can be significantly brought down around the phase transitions, which is consistent with the drop of multipartite entanglement depth seen in Fig.~\ref{fig:QFI_Ome8}.
In practice, we find that the angles determining the unitaries in Eq.~\eqref{eq:NL_NS} quickly converge with $\Omega$.

%%%%%%%%%%%%%%%%%%%%%%%%%%%%%%%%
%%%%%%%%%%%%%%%%%%%%%%%%%%%%%%%%
\subsection{Fermionic Non-Gaussianity}
\label{sec:non-gaussianity}
%%%%%%%%%%%%%%%%%%%%%%%%%%%%%%%%
%%%%%%%%%%%%%%%%%%%%%%%%%%%%%%%%
\noindent
The "magic" ingredient beyond entanglement can be investigated via other aspects than non-stabilizerness. For example, another resource that can enable potential quantum advantage is non-Gaussianity, relevant to fermionic or continuous bosonic quantum computing platforms~\cite{Terhal:2002,Bravyi:2005, Hebenstreit:2019, Dias:2024}. 
In fermionic systems, Gaussian states can be completely specified by their corresponding covariance matrix, or one-body correlation functions, 
and thus can be described efficiently classically. 
This motivates the measures of fermionic non-Gaussianity, also known as fermionic magic~\cite{Sierant:2025fax,Coffman:2025, Tarabunga:2026, Tarabunga:2026b}. 
Analogously to non-stabilizerness, the basic idea is to measure how far away the state is from the manifold of fermionic Gaussian states. The further away the state, the harder it is to approximate it efficiently classically and the higher is the non-Gaussianity. 

Deformed Hartree-Fock and BCS states are well-known examples of fermionic Gaussian states that can provide physical approximations to the Agassi ground state in the deformed and superfluid phase, respectively. Such states however explicitly break symmetries of the exact ground state, specifically the former lacks well-defined angular momentum, while the latter does not conserve particle number. The restoration of these symmetries subsequently takes the state away from the set of Gaussian states.

In this work we investigate the fermionic anti-flatness (FAF), which was recently introduced in Ref.~\cite{Sierant:2025fax} as measure of non-Gaussianity in fermionic systems.
\\

The FAF is based on the deviation of the covariance matrix $\Gamma$ of a state $\ket{\Psi}$ from that of a Gaussian state. Precisely, it measures the deviation of $\Gamma$ from the Gaussian orthogonality constraint $\Gamma^T \Gamma = \mathds{1}$, and is defined as~\cite{Sierant:2025fax}
\begin{align}
    \mathcal{F}_\eta (\ket{\Psi})=2\Omega-\frac{1}{2} \Tr \big[(\Gamma^T \Gamma)^\eta] \; .
\label{eq:FAF_def}
\end{align}
In Eq.~\eqref{eq:FAF_def} $\eta$ is a chosen integer, and $\Gamma_{ab}=-\frac{i}{2}\Braket{\Psi|[\gamma_a,\gamma_b]|\Psi}$ is the covariance matrix expressed in terms of the Majorana operators $\gamma_a$ with index $a=0,...,4\Omega-1$ in the doubled Majorana space~\footnote{The present Agassi model has $2\Omega$ fermionic orbitals.}
\begin{align}
    \gamma_{2m} = c_m + c_m^\dagger \; , \quad
    \gamma_{2m+1} = i (c_m - c_m^\dagger) \; .
\end{align}
where $m=0,...,2\Omega-1$ denotes fermionic orbitals.
Equivalently, we have, according to the JW mapping in Eq.~\eqref{eq:mapping}
\begin{align}
    \gamma_{2m}=\left(\prod_{l=0}^{m-1} (-Z_{l}) \right)X_m \; , 
    \quad 
    \gamma_{2m+1}=\left(\prod_{l=0}^{m-1} (-Z_l)\right)Y_m \; .
\end{align}
The covariance matrix $\Gamma_{ab}$ can then be expressed in terms of the one-body correlation matrix $\mathcal{C}_{mn} = \Braket{\Psi|c^\dagger_m c_n|\Psi}$ and pairing tensor $\kappa_{mn} = \Braket{\Psi|c_m c_n|\Psi}$. In a system with good particle number, such as the present Agassi ground state, the contribution of the latter cancels and the covariance matrix takes the form
\begin{align}
\Gamma_{ab} =
\begin{cases}
-i \mathcal{C}^-_{mn}  & \text{if } (a,b) = (2m,2n) \\ 
& \text{or } (a,b) = (2m+1,2n+1) \; , \\
\delta_{mn} - \mathcal{C}^+_{mn}  & \text{if } (a,b) = (2m,2n+1) \; , \\
- \left( \delta_{mn} - \mathcal{C}^+_{mn} \right) & \text{if } (a,b) = (2m+1,2n) \; .
\end{cases}
\end{align}
where $\mathcal{C}^\pm_{mn} \equiv \mathcal{C}_{mn} \pm \mathcal{C}_{nm}$.
Due to the symmetry of the Hamiltonian in Eq.~\eqref{eq:rescaled_Hami}, namely particle number, parity (and even parity per mode-pair), the one-body density matrix $\mathcal{C}_{mn} = n_m \delta_{mn}$ is diagonal, and thus the covariance matrix takes a very simple form in terms of occupation numbers $n_m$
\begin{align}
\Gamma_{ab} =
\begin{cases}
0  & \text{if } (a,b) = (2m,2n) \\ 
& \text{or } (a,b) = (2m+1,2n+1) \; , \\
\delta_{mn} (1- 2 n_m)  & \text{if } (a,b) = (2m,2n+1) \; , \\
- \delta_{mn} (1- 2 n_m) & \text{if } (a,b) = (2m+1,2n) \; .
\end{cases}
\label{eq:FAF_Agassi}
\end{align}

\begin{figure}[t]
    \centering
    \includegraphics[width=0.8\linewidth]{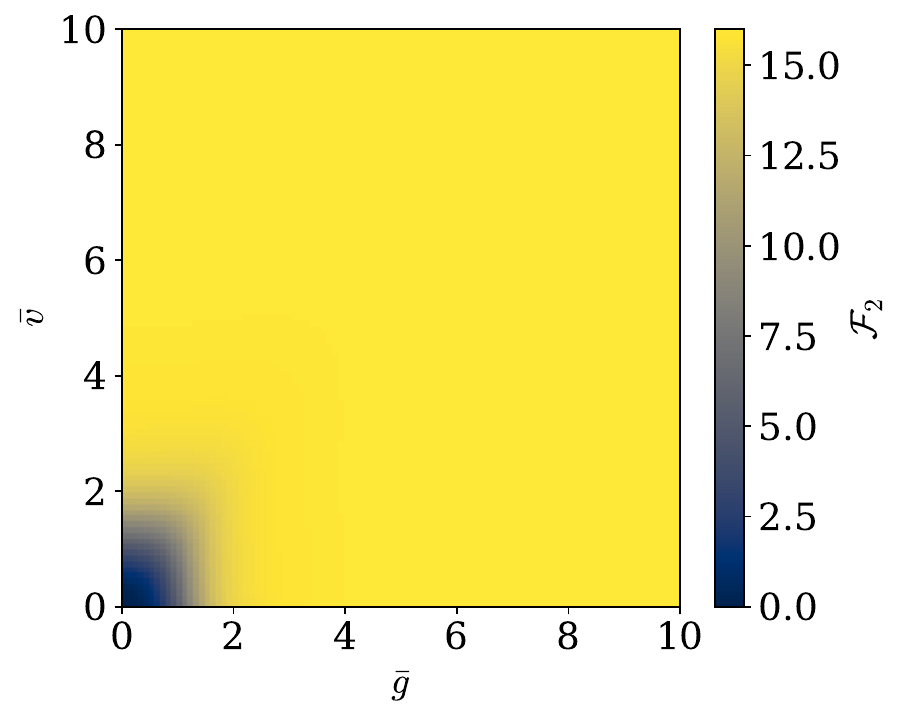}
    \caption{FAF $\mathcal{F}_2$ for $\Omega=8$ as a function of $\vbar$ and $\gbar$.}
    \label{fig:fantiflatness}
\end{figure}

Fig. \ref{fig:fantiflatness} displays the FAF of the Agassi ground state for $\eta=2$~\footnote{Similar behaviors are found for {\it e.g.} $\eta=1$.}. 
It is seen that the FAF is minimal for small $\vbar$ and $\gbar$ values, as the ground state is close to the simple non-interacting ground state, which is a Gaussian state.
As $\vbar$ and/or $\gbar$ increase, the FAF quickly saturates towards its maximum value, and exhibits an homogeneous behavior across the various regimes outside of the normal phase.
Since in the present model, the FAF reduces to the simple form in Eq.~\eqref{eq:FAF_Agassi}, 
it only captures leading-order correlation effects through occupation numbers, and thus is not sensitive to the nature of the correlations. 
\\ 

For comparison, we also computed the entropy of the closest gaussian state, a recently-introduced measure of non-Gaussianity from Ref.~\cite{Coffman:2025}.
We find this measure to behave in a very similar way as the FAF.
This is detailed in appendix \ref{app:rel_entropy}.

%%%%%%%%%%%%%%%%%%%%%%%%%%%%%%%%
%%%%%%%%%%%%%%%%%%%%%%%%%%%%%%%%
\section{Stabilizer ground states}
\label{sec:stabgs}
%%%%%%%%%%%%%%%%%%%%%%%%%%%%%%%%
%%%%%%%%%%%%%%%%%%%%%%%%%%%%%%%%
\noindent
Motivated by the behavior of the non-stabilizerness, and its non-local version, in the Agassi model, we explore to what extent the exact ground state can be captured by a (low-rank) superposition of stabilizer states, combined with rotations of the mode-pair basis.
%%%%%%%%%%%%%%%%%%%%%%%%%%%%%%%%
\subsection{Stabilizer Hamiltonians}
\label{sec:stab_Hami}
%%%%%%%%%%%%%%%%%%%%%%%%%%%%%%%%
In Ref.~\cite{Robin:2025wip} we develop a method for deriving stabilizer ground states given a Hamiltonian $\hat H$. The idea is to find an optimal division of $\hat H$ into a stabilizer part $\hat{H}_{\rm stab}$, and a residual non-stabilizer part $\hat W$, as
\begin{align}
\hat H &= \hat{H}_{\rm stab} + \hat W \nonumber \\
       &= \sum_{P \in \mathcal{S}} a_P \hat P  +  \sum_{P \in \mathcal{G}_{N_q}(H) \notin \mathcal{S}} a_P \hat P \; , 
\label{eq:Hami_sep_gen}
\end{align}
where $\mathcal{S}$ forms a (complete) stabilizer group with $2^{N_q}$ commuting elements (some of the coefficients $a_P$ may be equal to zero). 
Ideally, the stabilizer Hamiltonian $\hat{H}_{\rm stab}$ should capture as much physics as possible, so that it provides a good approximation $\ket{\Psi_s}$ to the exact ground state. In a second step, the non-stabilizer term $\hat W$ can be treated via perturbative methods if it is "small" enough, or using powerful classical or, at scale, quantum algorithms such as, {\it e.g.} (quantum) imaginary time evolution (ITE)~\cite{Motta:2019} or quantum phase estimation~\cite{Kitaev:1995}. 
In Ref.~\cite{Robin:2025wip} this method was applied to the LMG model, for which the stabilizer Hamiltonian $H_{\rm stab}$ was determined using an energy-minimization criterion. The corresponding stabilizer ground state was found to provide a high-quality approximate ground state in the parity-broken (collective) phase, which led to acceleration of the convergence of ITE.
In the present Agassi model, given the behavior of the SRE $\M_2$, one can expect that a single stabilizer state may also be able to capture the full deformed phase ($\vbar \gtrsim \gbar$) to a large extent. The superfluid phase, however, as well as the phase transitions, present significant non-stabilizerness and thus are expected to require superpositions of a number of stabilizer states. 
Here we investigate if a low-rank superposition is able to provide a reasonable approximation to the ground state in these regimes. We also explore how rotations of the mode-pairs impact the fidelity and rank of this stabilizer superposition.

We follow the technique form Ref.~\cite{Robin:2025wip} to determine the stabilizer Hamiltonians, and corresponding ground states. That is, given the Hamiltonian in Eq.~\eqref{eq:rescaled_Hami}, we further separate the pairing term into diagonal and off-diagonal terms
\begin{align}
    \widetilde{H} &= \widetilde{H}_0 + \widetilde{H}_{v} +  \widetilde{H}_g \nonumber \\
    &= \widetilde{H}_0 + \widetilde{H}_{v} +  \widetilde{H}_g^{diag} + \widetilde{H}_g^{ndia, I} + \widetilde{H}_g^{ndia, II} \; ,
\label{eq:Hami_rescaled_II}
\end{align}
where the various terms are detailed in appendix~\ref{app:Hamiltonian}.
Instead of looking for optimal stabilizer groups in a "blind" way, we use the natural separation in Eq.~\eqref{eq:Hami_rescaled_II}, and look for stabilizer groups within these terms.
As different parts of the Hamiltonian dominate depending on the choice of $\vbar$ and $\gbar$, different stabilizer Hamiltonians $\hat{H}_{stab}$ will emerge in the various regions of the phase diagram.
Step-by-step procedures for low values of $\Omega$ are provided in appendix~\ref{app:stabsforsmall}.
Below we provide results for general values of $\Omega$.

 \begin{itemize}
    \item The single-particle term $\widetilde{H}_0$ yields a stabilizer group $\mathcal{S}_1 = \langle +Z_0, - Z_1, +Z_2, -Z_3, ...., -Z_{2\Omega-1} \rangle$ with total stabilizer energy $E(\mathcal{S}_1) = -\frac{\Omega}{2}\left(1+\frac{1}{\Omega-1}\gbar\right)$ and $\ket{\Psi}_{s,1} = \ket{01...01}$, which is the non-interacting ground state, where all particles are on the lower level.
    \item Terms from $\widetilde{H}_{v}$ yield $\mathcal{S}_2 = \langle X_0X_1X_2X_3, Y_0Y_1Y_2Y_3,$ $ X_0X_1Y_2Y_3, X_0X_1X_4X_5, X_0X_1Y_4Y_5, ...,$ $ X_0X_1X_{2\Omega-2}X_{2\Omega-1}, X_0X_1Y_{2\Omega-2}Y_{2\Omega-1},$ $ X_0 Y_1 Y_2 X_3 ... X_{2\Omega-4} Y_{2\Omega-3} Y_{2\Omega-2} X_{2\Omega-1}\rangle$ with energy $E(\mathcal{S}_2)  = -\frac{\Omega}{2}\left(\frac{1}{\Omega-1}\gbar+\frac{1}{2}\vbar\right)$.
    The corresponding stabilizer state $\ket{\Psi}_{s,2}$ is an equal superposition of the non-interacting ground state and all possible fermionic excitations that can be generated through $\widetilde{H}_{v}$.
    \item Terms common to $\widetilde{H}_{v}$ and $\widetilde{H}_{g}^{ndiag,I}$ provide a stabilizer group $\mathcal{S}_3 = \langle X_0X_1X_2X_3, Y_0Y_1Y_2Y_3, X_0X_1Y_2Y_3, X_0Y_1Y_2X_3,...,$ $ X_{2\Omega-4}X_{2\Omega-3}X_{2\Omega-2}X_{2\Omega-1}, Y_{2\Omega-4}Y_{2\Omega-3}Y_{2\Omega-2}Y_{2\Omega-1},$ $ X_{2\Omega-4}X_{2\Omega-3}Y_{2\Omega-2}Y_{2\Omega-1}, X_{2\Omega-4}Y_{2\Omega-3}Y_{2\Omega-2}X_{2\Omega-1} \rangle$ with $E(\mathcal{S}_3)  = -\frac{\Omega}{\Omega-1}\left(\gbar+\frac{1}{2}\vbar\right)$, and
    \begin{align}
        \ket{\Psi}_{s,3} = -  \left(\frac{1}{\sqrt{2}}\ket{1010} + \ket{0101} \right)^{\otimes{\Omega/2}} \; ,
    \end{align}
    which is a tensor product of entangled mode-pairs.
    \item The Hamiltonian term $\widetilde{H}_{g}^{ndiag,II}$ yields several degenerate stabilizer groups.
    %Here, we do not only find one stabilizer ground state but several degenerate states. Every stabilizer group 
    Each of them consists of $\Omega/2$  sets $s_{i,j}$, where $i+3\leq j\in \{0,1,4,5,...,(2\Omega-4)(2\Omega-3)\}$ and 
    \begin{align}
        s_{i,j} \equiv \{ X_i Z_{i+1} X_{i+2} X_j Z_{j+1} X_{j+2},\nonumber \\
        Y_i Z_{i+1} Y_{i+2} Y_j Z_{j+1} Y_{j+2}, \nonumber \\
        -X_i Z_{i+1} X_{i+2} Y_j Z_{j+1} Y_{j+2},\nonumber \\ 
        X_i Z_{i+1} Y_{i+2} Y_j Z_{j+1} X_{j+2} \}
    \end{align} 
    with $E(\mathcal{S}_4^{(a)} ) = -\frac{\Omega}{\Omega-1}\gbar$, with $a=1,...,N_g$.
    There are $N_g(\Omega)=\frac{1}{2}\binom{\Omega}{\Omega/2}-\frac{\Omega}{2}=\binom{\Omega-1}{\Omega/2-1}-\frac{\Omega}{2}$, for $\Omega>4$, corresponding stabilizer states, each of them containing pairs excitations compatible with $\widetilde{H}_{g}$.
    \item The diagonal pairing term $\widetilde{H}_{g}^{diag}$ provides $\mathcal{S}_5 = \langle +Z_0, + Z_1, +Z_2, +Z_3, ... +Z_{2\Omega-1} \rangle$ with energy $E(\mathcal{S}_5)  = -\frac{\Omega}{\Omega-1}\gbar$. However the corresponding stabilizer state $\ket{\Psi}_{s,5} = \ket{0...0}$ violates particle number conservation, and therefore we discard this group.
\end{itemize}

%%%%%%%%%%%%%%%%%%%%%%%%%%%%%%%%%%%%%%%%%%%
\subsection{Preparing Stabilizer States from Graph States}
\begin{figure}[h]
    \centering
    \subfloat[]{\includegraphics[width=0.33\linewidth]{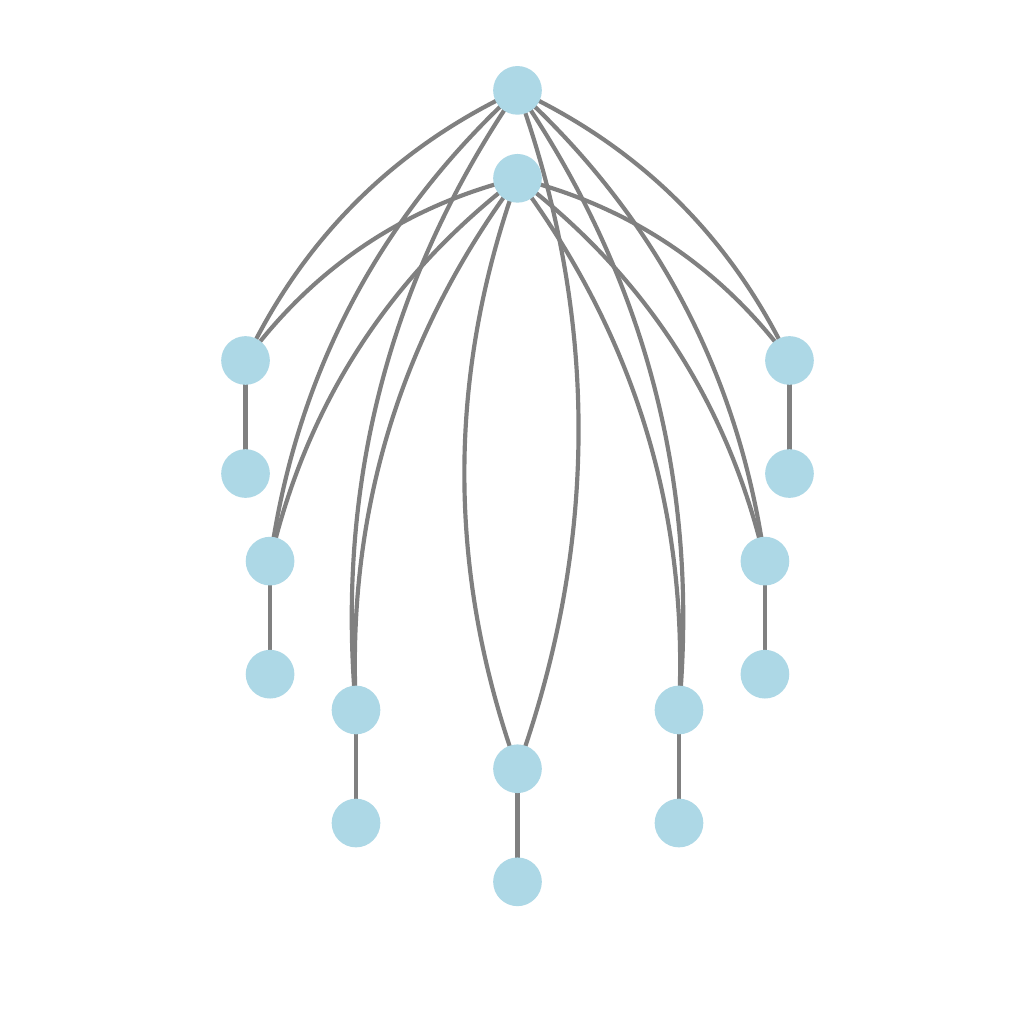}}
    \hfill
    \subfloat[]{\includegraphics[width=0.33\linewidth]{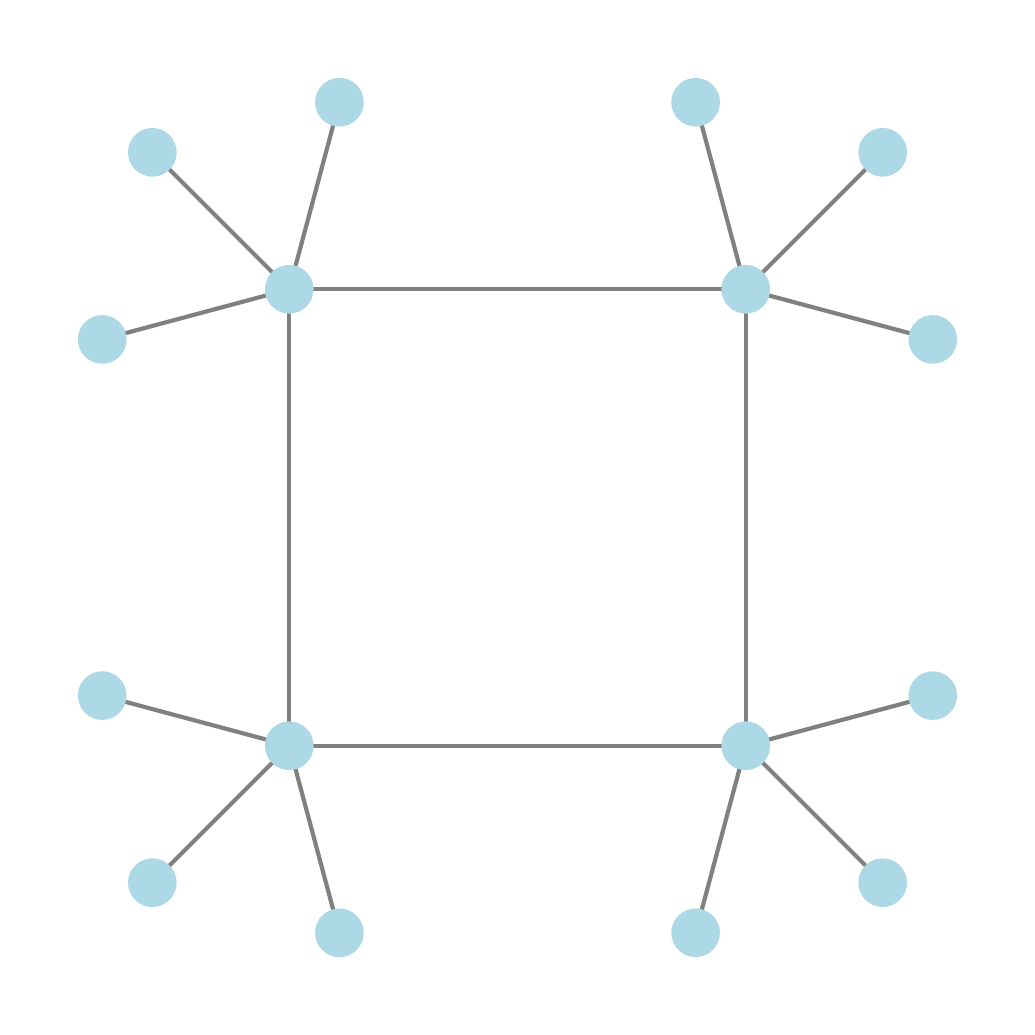}}
    \hfill
    \subfloat[]{\includegraphics[width=0.33\linewidth]{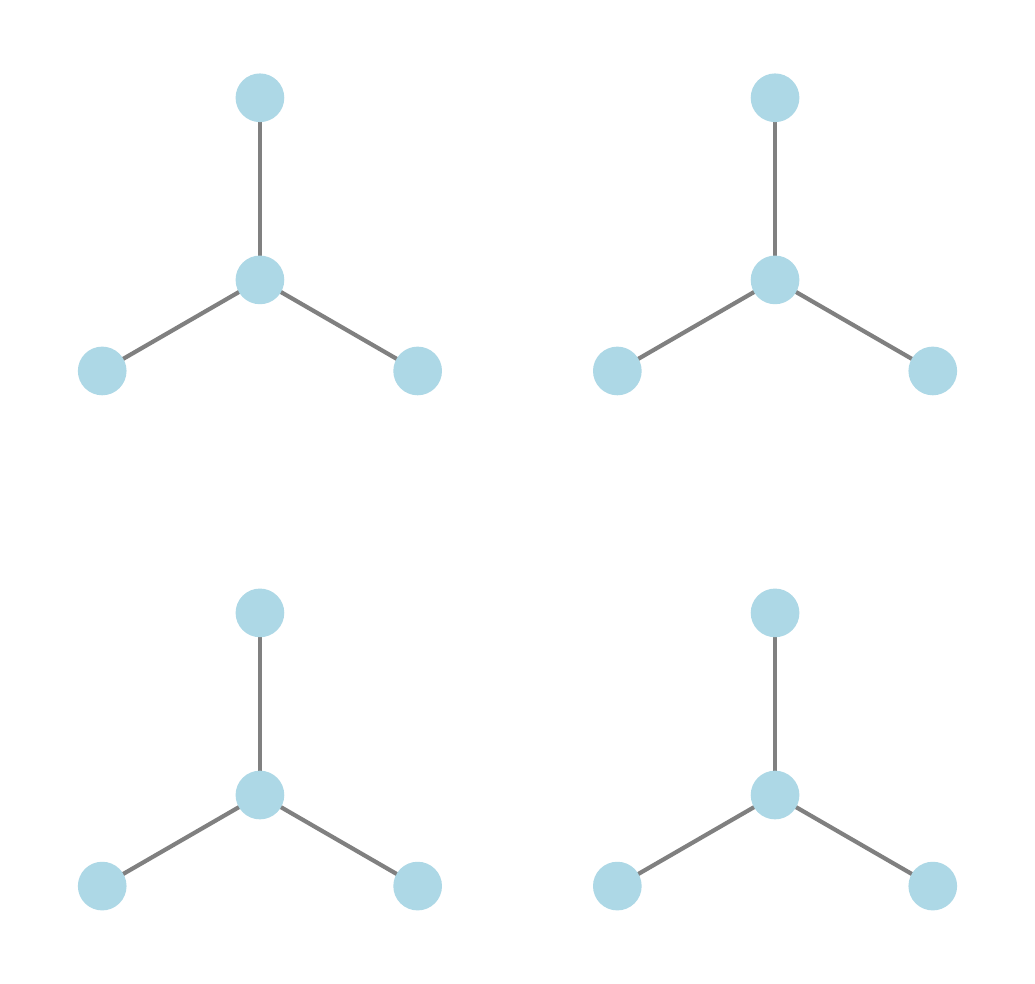}}
\caption{(a) Graph state for the LMG phase (LC-equivalent to state stabilized by group $\mathcal{S}_2$); 
and (b) \& (c) Graph states for the pairing phase (LC-equivalent to state stabilized by group of the type $\mathcal{S}_4$), the graph states are connected or disconnected depending on the specific group.}
\label{fig:graphs}
\end{figure}

%%%%%%%%%%%%%%%%%%%%%%%%%%%%%%%%%%%%%%%%%%%
\label{sec:stabsfromgraph}
Each of the stabilizer states above can be prepared efficiently. To do that, we follow the same procedure as in Ref.~\cite{Robin:2025wip}, which consists in first preparing the local-Clifford-equivalent (LC-equivalent) graph state, followed by application of local Clifford operations to generate the targeted stabilizer state. This can be done following, for example, the procedure from Ref.~\cite{Nest:2004khg} by performing a reduction of the stabilizer tableau (keeping track of the generator signs) to obtain the adjacency matrix of the graph.
This was described in Ref.~\cite{Robin:2025wip} and applied to the LMG model in the spin formulation.
As examples, we show in Fig. \ref{fig:graphs} the graph state corresponding to the stabilizer groups $\mathcal{S}_2$, characteristic of the LMG regime ($\vbar>2$, $\gbar=0$)~\footnote{Graph~\ref{fig:graphs}(a) with the present orbital-to-qubit mapping is the analogous of the graph in Ref.~\cite{Robin:2025wip} which used a spin-to-qubit mapping of the LMG model.}, and some graph states corresponding to stabilizer groups $\mathcal{S}_4^{(a)}$, characteristic of the superfluid regime ($\vbar=0$, $\gbar>2$), for $\Omega=8$.

%%%%%%%%%%%%%%%%%%%%%%%%%%%%%%%%%%%%%%%%%%%
\subsection{Stabilizer-State Expansions}
%%%%%%%%%%%%%%%%%%%%%%%%%%%%%%%%%%%%%%%%%%%
\label{sec:stabexpansion}

As an ansatz to describe the ground state approximation, we choose a superposition of the stabilizer states $\ket{\Psi}_{s,i}$ from Sec.~\ref{sec:stab_Hami}
\begin{align}
    \ket{\Psi}_{r} &= \sum_{i=1}^r  c_i \ket{\Psi}_{s,i} \nonumber \\
    & = c_1 \ket{\Psi}_{s,1} + c_2 \ket{\Psi}_{s,2} + c_3 \ket{\Psi}_{s,3} 
    + \sum_{a=1}^{N_g(\Omega)} c_{4,a} \ket{\Psi}_{s,4}^{(a)} \; ,
\label{eq:LRS}
\end{align}
and study how the stabilizer rank $r$ varies in different regimes, for a given required precision on the energy, or fidelity. 

To find the optimal ground-state approximation, the coefficients $c_1$, $c_2$, $c_3$ and $c_{4,a}$ are optimized such that the energy $E_r = \langle \Psi_{r} |\widetilde{H}| \Psi_{r} \rangle$ is minimal. The left side of Fig.~\ref{fig:rotated_LRS} shows the fidelity and relative energy difference of the resulting stabilizer superposition $\ket{\Psi}_{r}$, relative to the exact ground state, using the maximal rank value $r=N_g(\Omega) +3$. The stabilizer expansion is able to capture all three phases to almost $100\%$ fidelity, however the quality degrades near the phase transitions, especially for small $\vbar$ and $\gbar\approx1-2$, where the fidelity drops to $\approx 0.77$. 
Similarly the relative energy error becomes larger where the fidelity decreases.

The convergence of the stabilizer expansion as we increase the rank value from $r=1$ to $r=N_g(\Omega)+3$ is shown in appendix~\ref{app:LRS_trunc}.
While the spherical and deformed phase are largely described with a single stabilizer state ($\ket{\Psi}_{s,1}$ and $\ket{\Psi}_{s,2}$, respectively), the stabilizer expansion in the superfluid phase converges much more slowly, and requires an exponentially-growing number of stabilizer states ($\propto 2^\Omega / \sqrt{\Omega}$) to reach a given fidelity~\footnote{
As discussed in Sec. \ref{sec:stab_Hami}, the number of degenerate stabilizer ground states in the superfluid phase is $N_g(\Omega)=\frac{1}{2}\binom{\Omega}{\Omega/2}-\frac{\Omega}{2}=\binom{\Omega-1}{\Omega/2-1}-\frac{\Omega}{2}$, for $\Omega>4$. Using Stirling's formula, this corresponds to an exponential scaling $N_g(\Omega)\sim \frac{2^\Omega}{\sqrt{2\pi \Omega}}$. }. This number nevertheless remains exponentially smaller than the dimension of the half-filled, even-parity computational-basis sector (which scales as $\propto 4^\Omega / \sqrt{\Omega}$).
%

%%%%%%%%%%%%%%%%%%%%%%%%%%%%%%%%%%%%%%%%%%%
\subsection{Basis-Rotated Stabilizer Expansions}
%%%%%%%%%%%%%%%%%%%%%%%%%%%%%%%%%%%%%%%%%%%
\label{sec:br-stabexp}

As a last step, we explore whether basis optimization may allow for more efficient truncations in the stabilizer expansion, in the superfluid regime, and around critical points.
This is motivated by the behavior of the non-stabilizerness, which contains a significant local component around the phase transitions, which can thus be largely eliminated via mode-pair rotations, as shown in section~\ref{sec:NLmagic}.

To perform the basis-rotated stabilizer expansion of the Agassi ground state, there are in principle two approaches that one can take:
\begin{enumerate}
    \item i) Unitarily transform the Hamiltonian through $\hat{H}_{\rm eff} = \hat{U}^\dagger \hat{H} \hat{U}$, where $\hat{U} = \hat{U}_{A_1} \otimes ... \otimes \hat{U}_{A_{\Omega/2}}$ is a tensor product of unitaries acting on the ${\Omega/2}$ mode-pairs, constrained to preserve the relevant symmetries, as detailed in appendix~\ref{app:NL_magic}. ii) Expand the resulting effective Hamiltonian $\hat{H}_{\rm eff}$ in terms of Pauli matrices, search for stabilizer groups in the same way as for $\hat{H}$, to which correspond stabilizer states that can be used to expand the ground state. 
    \item Directly rotate the stabilizer states from Sec.~\ref{sec:stab_Hami} as $\ket{{\Psi'}}_{s,i} = \hat{U} \ket{\Psi}_{s,i}$.
\end{enumerate}
While the first approach is more general,  
optimizing over the stabilizer groups may become increasingly hard for large values of $\Omega$.
For that reason we will adopt here the second approach, which can be handled easily.
In this case we do not look for new stabilizer groups, however the rotated $\ket{{\Psi'}}_s$ are not stabilizer states in the original basis, they are a superposition of states with support outside the previous configuration space of $\ket{\Psi}_r$ in Eq.~\eqref{eq:LRS}.
The basis rotation $\hat{U} = \hat{U}_{A_1} \otimes ... \otimes \hat{U}_{A_{\Omega/2}}$ is optimized to minimize the energy of the rotated stabilizer expansion $\ket{\Psi'} = \sum_{i=1}^r \ket{{\Psi'}}_{s,i}$.

\begin{figure}[h]
    \centering
    \subfloat{\includegraphics[width=0.49\linewidth]{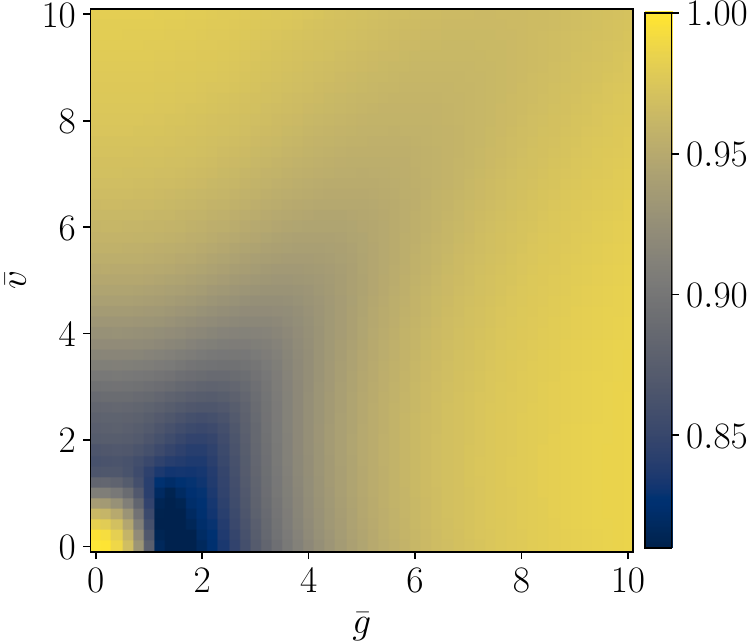}}
    \hfill
    \subfloat{\includegraphics[width=0.49\linewidth]{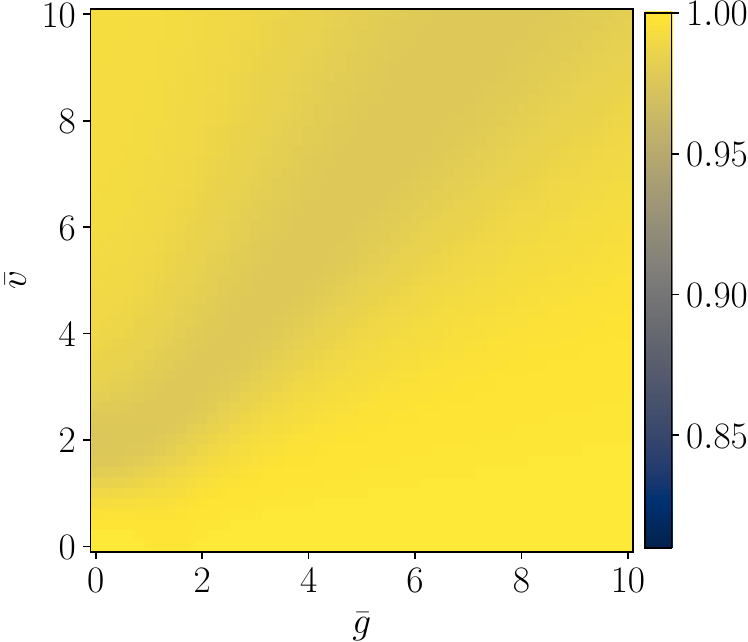}}
    \hfill
    \subfloat{\includegraphics[width=0.49\linewidth]{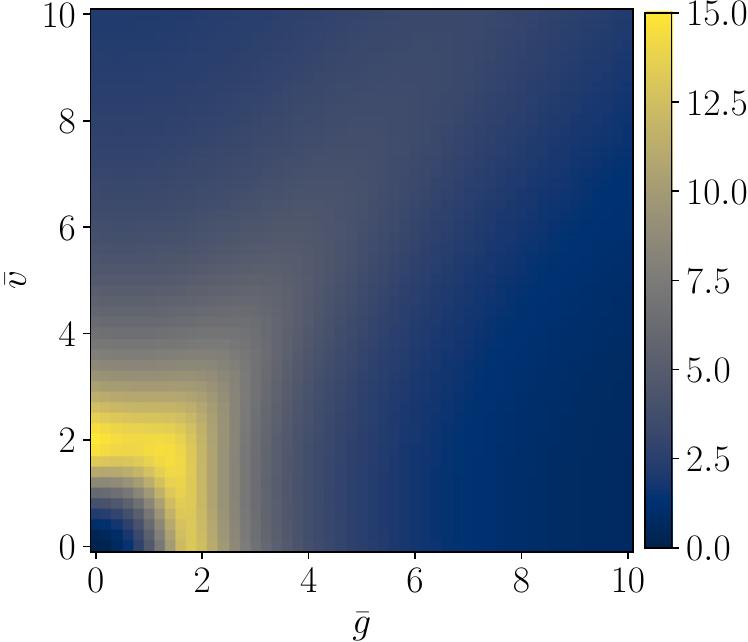}}
    \hfill
    \subfloat{\includegraphics[width=0.49\linewidth]{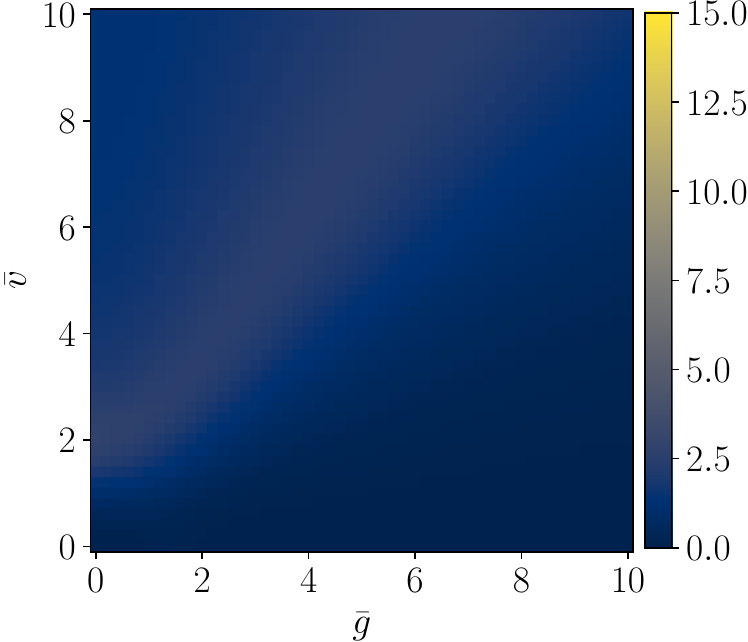}}
\caption{
Upper panels: Fidelity between the bare (left) and rotated (right) superposition of stabilizer states and the exact ground state for $\Omega=8$. 
Lower panels: Same for the relative energy difference in $\%$.
}
\label{fig:rotated_LRS}
\end{figure}

The fidelity and relative energy difference of the resulting stabilizer expansion $\ket{\Psi'}$ are shown in Fig.~\ref{fig:rotated_LRS} on the right. For comparison, the fidelity without basis optimization is displayed in the left panels. 
The basis rotation substantially improves the results and yields highly-accurate wavefunctions, with fidelity above $ 97 \% $ and energy accuracy below $3\%$, across the entire phase diagram.
Furthermore, in the regimes that are harder to capture with stabilizer expansions, where the pairing interaction is significant, wavefunctions with large fidelities can be reached with significantly less stabilizer states than without basis rotations. This is illustrated in Fig.~\ref{fig:trunc_LRS} of App.~\ref{app:LRS_trunc}. Such states can potentially be used as initial states for refinement with quantum algorithms such as, {\it e.g.} quantum phase estimation or imaginary time evolution.

%%%%%%%%%%%%%%%%%%%%%%%%%%%%%%%%%%%%%%%%%%%%%%%%%%%%
%%%%%%%%%%%%%%%%%%%%%%%%%%%%%%%%%%%%%%%%%%%%%%%%%%%%
\section{Summary and Conclusions}
\label{sec:conclu}
%%%%%%%%%%%%%%%%%%%%%%%%%%%%%%%%%%%%%%%%%%%%%%%%%%%%
%%%%%%%%%%%%%%%%%%%%%%%%%%%%%%%%%%%%%%%%%%%%%%%%%%%%

In this work, we studied aspects of many-body quantum complexity in the ground-state of the Agassi Hamiltonian in terms of entanglement, (non-local) non-stabilizerness, and non-gaussianity. This model, which possesses a rich phase diagram, allows us to investigate how these aspects of complexity relate to the emergence of collectivity—including deformation, superfluid pairing, the interplay between them, and their coupling to single-particle degrees of freedom.

While the highly-entangled, yet nearly stabilizer character, remains over the deformed phase (regime with small pairing interaction), the origin of the simplicity observed in the superfluid pairing phase 
does not appear to reside in a single low measure of quantum complexity (among those studied).
In particular, both multipartite entanglement and non-stabilizerness appear to be significant, while FAF (characterizing fermionic non-gaussianity) is found to be maximal.
Thus, unlike in the deformed phase, the underlying microscopic mechanism for compression of the many-body state in the pairing regime remains unclear from the perspective of quantum information and quantum resources.
We also observe that while non-stabilizerness peaks at phase transitions, a significant amount can be removed via rotations of the "natural" local degrees-of-freedom (mode-pairs). Conversely, the multipartite character decreases at the phase transition. 
Together, these trends suggest that the transition involves a reorganization of the ground state: multipartite correlations weaken as the system shifts to a new natural mode-pair basis, then build up again within the new phase.

We then exploited these findings to describe the ground state of the system as a superposition of stabilizer ground states, based on a decomposition of the Hamiltonian into stabilizer parts, where each stabilizer sector is determined by the physics of the system.
Consistent with the observed measures of non-stabilizerness, we found that a single stabilizer state is sufficient to capture the entire deformed phase to a large extent, whereas an accurate description of the pairing phase requires a number of degenerate stabilizer ground states that grow exponentially with system size (although exponentially fewer than the computational-basis states in the same symmetry sector).
We found however, that, for the systems studied, local basis rotations selected through energy minimization can yield high-fidelity approximations to the exact ground state, through the entire phase diagram,  using a number of stabilizer states of the order of the system size.
This, in a way, combines stabilizer ground state methods with variational basis optimizations implemented, for example, in Refs.~\cite{Robin:2023,Hengstenberg:2023}.
The resulting ground-state approximations can then serve as starting point for quantum algorithms such as, {\it e.g.} quantum phase estimation. 
The next step is to extend this work to more complex and realistic systems, such as nuclear ground states.

%%%%%%%%%%%%%%%%%%%%%%%%%%%%%%%%
\begin{acknowledgements}
We would like to thank Martin J. Savage for useful discussions. We also acknowledge interesting discussions held during the second workshop on Many-Body Quantum Magic held at the InQubator for Quantum Simulation at the University of Washington in Seattle (2025).
This work was supported, in part, by Universit\"at Bielefeld, and by the Deutsche Forschungsgemeinschaft (DFG, German Research Foundation) through the CRC-TR 211 'Strong-interaction matter under extreme conditions'– project number 315477589 – TRR 211.
\end{acknowledgements}

\bibliography{biblio_paper}

\onecolumngrid

%%%%%%%%%%%%%%%%%%%%%%%%%%%%%%%%%%%%%%%%%%%%%%%%%%%%
%%%%%%%%%%%%%%%%%%%%%%%%%%%%%%%%%%%%%%%%%%%%%%%%%%%%
\appendix
%%%%%%%%%%%%%%%%%%%%%%%%%%%%%%%%%%%%%%%%%%%%%%%%%%%%
%%%%%%%%%%%%%%%%%%%%%%%%%%%%%%%%%%%%%%%%%%%%%%%%%%%%

%%%%%%%%%%%%%%%%%%%%%%%%%%%%%%%%%%%%%%%%%%%%%%%%%%%%
\section{The Agassi Hamiltonian mapped onto qubits}
\label{app:Hamiltonian}
%%%%%%%%%%%%%%%%%%%%%%%%%%%%%%%%%%%%%%%%%%%%%%%%%%%%
The Hamiltonian of the Agassi model, Eq. \eqref{eq:agassiDEF}, can be rewritten in terms of Pauli strings using the pa-JW mapping in Eq.~\eqref{eq:mapping}. 
The (rescaled) Hamiltonian then writes
\begin{equation}
\frac{H}{\varepsilon} \equiv \widetilde{H} = \widetilde{H}_0 + \widetilde{H}_{v} +  \widetilde{H}_{g} \; ,
\label{eq:QHam}
\end{equation}
where the single-particle term is given by
\begin{align}
\widetilde{H}_0 = \frac{1}{4} \sum_{m= 0}^{2\Omega-1} (-1)^{m+1} Z_m \; ,
\end{align}
and the particle-hole, or LMG, interaction writes
\begin{align}
\widetilde{H}_v = - \frac{\vbar}{\Omega-1}  \sum_{m,n \, odd  \atop(m<n)} Q_{m-1,m,n-1,n} \; ,
\end{align}
with
\begin{align}
Q_{m-1,m,n-1,n} 
 \equiv  &\left( \sigma_{m-1}^- \sigma_m^+ \sigma_{n-1}^- \sigma_n^+ + h.c. \right) \nonumber \\
 = &\frac{1}{8} \bigl( X_{m-1} X_m X_{n-1} X_n + X_{m-1} X_m Y_{n-1} Y_n  
 - X_{m-1} Y_m X_{n-1} Y_n  + X_{m-1} Y_m Y_{n-1} X_n  \nonumber \\
& + Y_{m-1} X_m X_{n-1} Y_n  - Y_{m-1} X_m Y_{n-1} X_n  + Y_{m-1} Y_m X_{n-1} X_n  + Y_{m-1} Y_m Y_{n-1} Y_n  \bigr) \; .
\end{align}
The pairing term contains three types of contributions
\begin{align}
\widetilde{H}_g = \widetilde{H}_g^{diag} + \widetilde{H}_g^{ndia, I} + \widetilde{H}_g^{ndia, II} \; ,
\end{align}
where the diagonal term is
\begin{align}
\widetilde{H}_g^{diag} = -\frac{\gbar}{\Omega-1} \left( \sum_{m=0,4,8...} +  \sum_{m=1,5,9...} \right) \Lambda_m \Lambda_{m+2} \; ,
\end{align}
with $\Lambda_m = (\mathds{1}_m + Z_m)/2$,
and the non diagonal terms are
\begin{align}
\widetilde{H}_g^{ndia, I} = -\frac{\gbar}{\Omega-1} \sum_{m=0,4,8...} Q_{m,m+1,m+2,m+3} \; ,
\end{align}
and 
\begin{align}
\widetilde{H}_g^{ndia, II} =& -\frac{\gbar}{\Omega-1} \bigg( \sum_{m,n=0,4,8... \atop (m<n)} + \sum_{m,n=1,5,9...\atop (m<n)} 
+ \sum_{m=0,4,8..; n=1,5,9... \atop (|n-m|>1)} \bigg) \overline{Q}_{m,m+1,m+2;n,n+1,n+2} \; ,
\end{align}
where 
\begin{align}
\overline{Q}_{m,m+1,m+2;n,n+1,n+2} 
 \equiv& \big( \sigma_{m}^+ Z_{m+1} \sigma_{m+2}^+ \sigma_{n}^- Z_{n+1} \sigma_{n+2}^- + h.c. \big) \nonumber \\
 =& \frac{1}{8} \bigl( X_{m} Z_{m+1} X_{m+2} X_{n} Z_{n+1} X_{n+2} 
 - X_{m} Z_{m+1} X_{m+2} Y_{n} Z_{n+1} Y_{n+2}  \nonumber \\
& + X_{m} Z_{m+1} Y_{m+2} X_{n} Z_{n+1} Y_{n+2}  
+ X_{m} Z_{m+1} Y_{m+2} Y_{n} Z_{n+1} X_{n+2}  \nonumber \\
&+ Y_{m} Z_{m+1} X_{m+2} X_{n} Z_{n+1} Y_{n+2} 
+ Y_{m} Z_{m+1} X_{m+2} Y_{n} Z_{n+1} X_{n+2}  \nonumber \\
&- Y_{m} Z_{m+1} Y_{m+2} X_{n} Z_{n+1} X_{n+2}  
+ Y_{m} Z_{m+1} Y_{m+2} Y_{n} Z_{n+1} Y_{n+2}  \bigr) \; .
\end{align}
%

%%%%%%%%%%%%%%%%%%%%%%%%%%%%%%%%%%%%%%%%%%%%%%%%%%%%
\section{Entanglement in the Agassi ground state}
\label{app:entanglement}
\begin{figure}[!h]
\centering{\includegraphics[width=0.45\columnwidth] {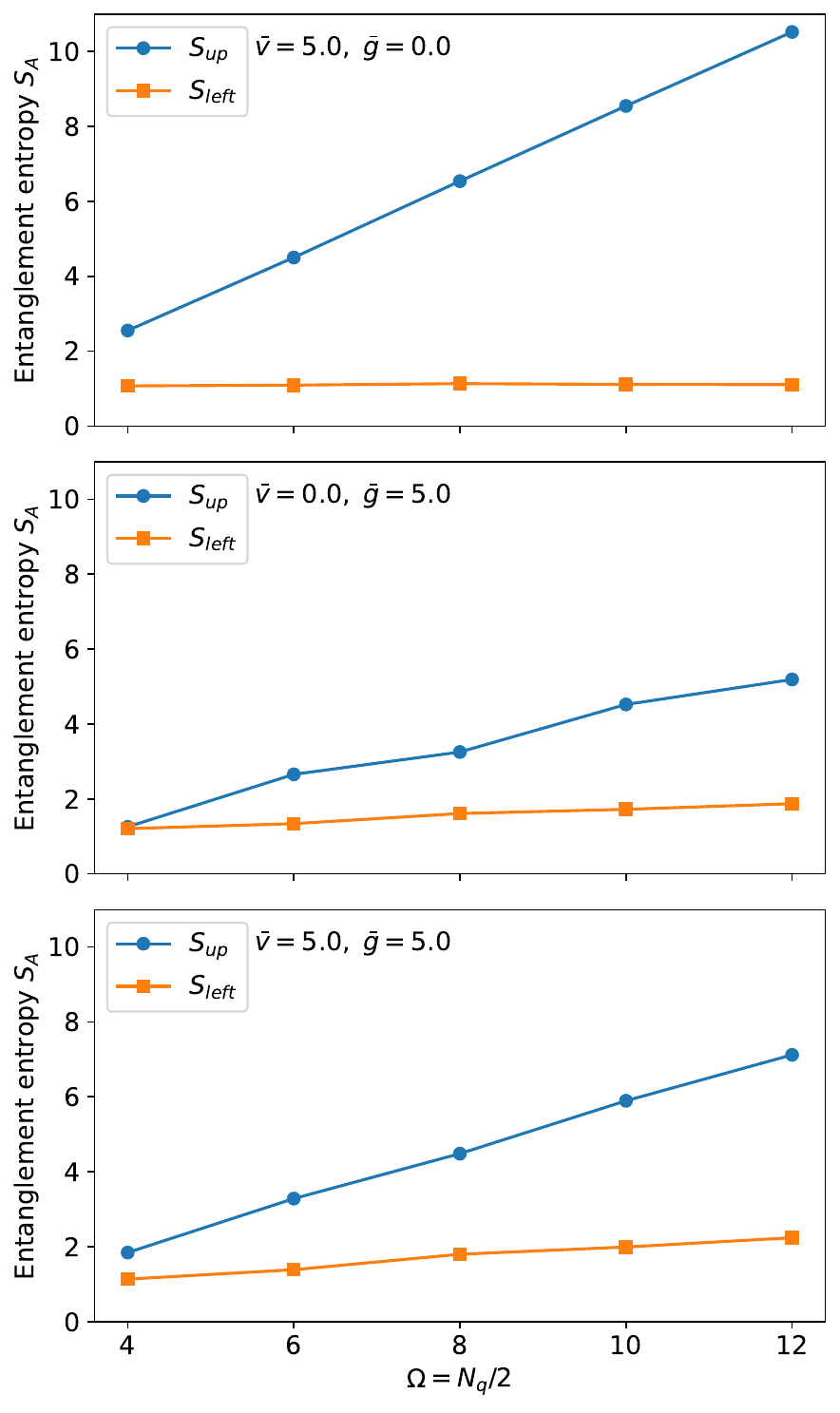}} 
\caption{Entanglement entropies $S_A$ as a function of system size $\Omega = N_q/2$, for the two bi-partitionings of the system: up/down (blue) and left/right (orange). We study three limits $(\vbar,\gbar)=(5.0,0.0), \; (0.0, 5.0), \; (5.0, 5.0)$ displayed on the upper, middle and lower panels, respectively.
The upper and middle panel correspond to the LMG and pairing limit in the deformed and superfluid phases, respectively, while the lower panel corresponds to a critical region of the Agassi model at the transition between these two phases.
}
\label{fig:vNentropy_scaling}
\end{figure}

%%%%%%%%%%%%%%%%%%%%%%%%%%%%%%%%%%%%%%%%%%%%%%%%%%%%
We show in Fig.~\ref{fig:vNentropy_scaling} the entanglement entropies for the two bi-partitionings studied in section~\ref{sec:entang} as a function of system size $\Omega=N_q/2$.
%

%%%%%%%%%%%%%%%%%%%%%%%%%%%%%%%%%%%%%%%%%%%%%%%%%%%%
\section{Non-Stabilizerness in the Agassi ground state}
\label{app:NS}
\begin{figure}[h!]
    \centering
    \subfloat[$\gbar=0$]{\includegraphics[width=0.49\textwidth]{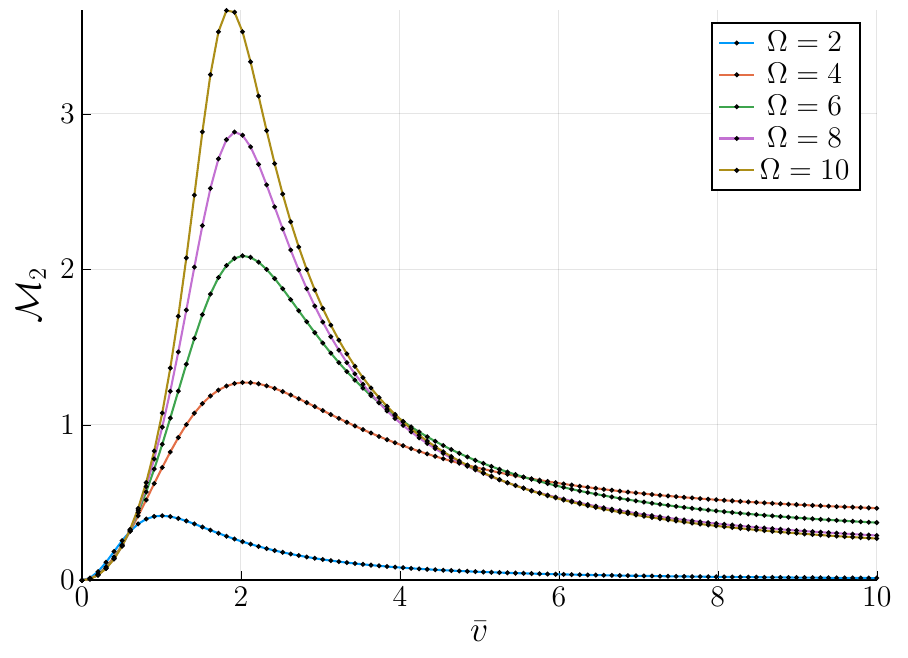}}
     \subfloat[$\vbar=0$]{\includegraphics[width=0.49\textwidth]{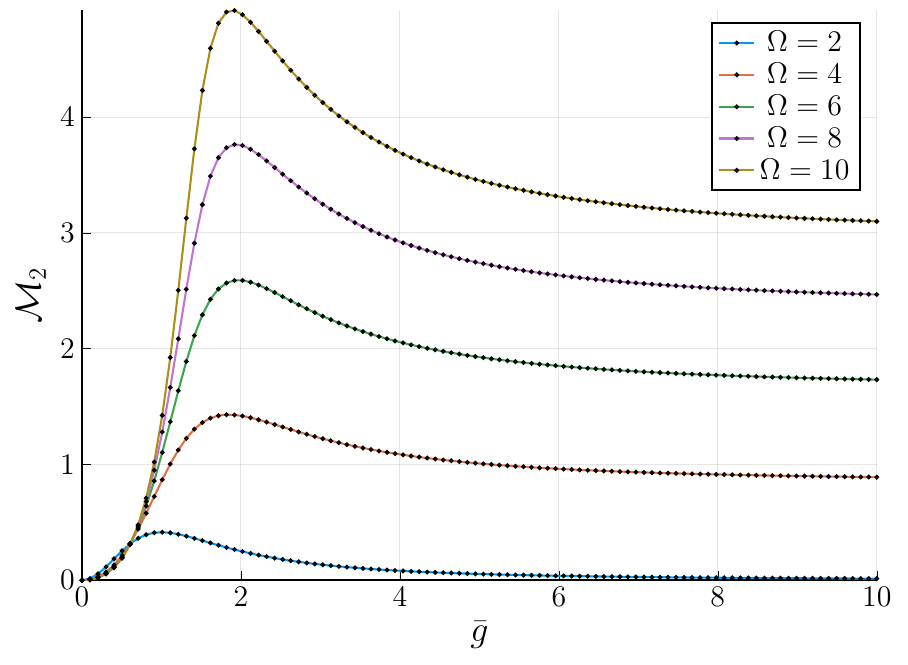}}
     \hfill
     \subfloat[$M_2$ along straight line from $(0,0)$ to $(10,9)$.]{\includegraphics[width=0.49\textwidth]{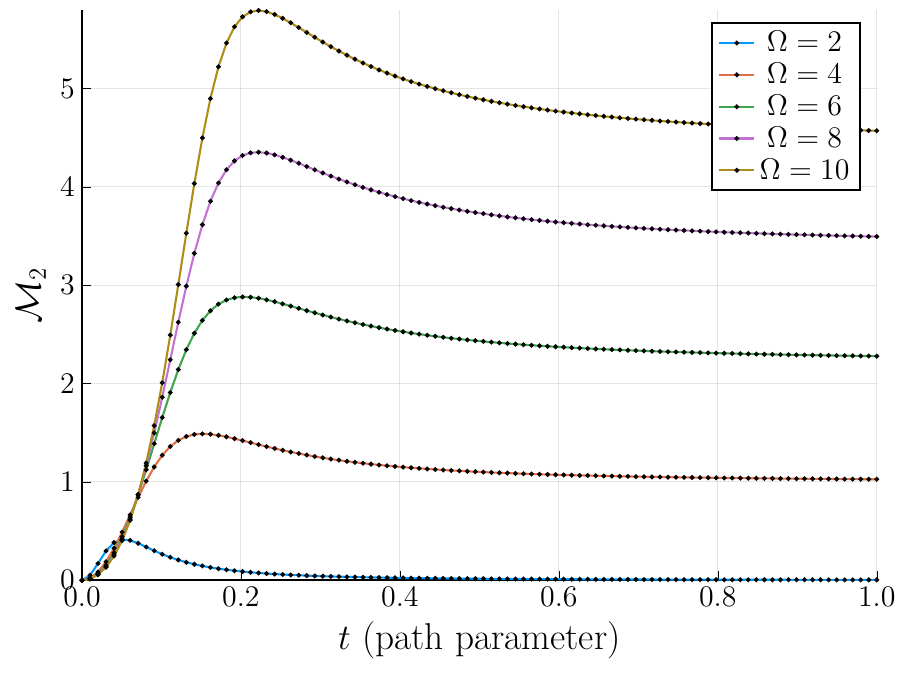}}
     \subfloat[$\vbar=0$, $\gbar=10$]{\includegraphics[width=0.49\textwidth]{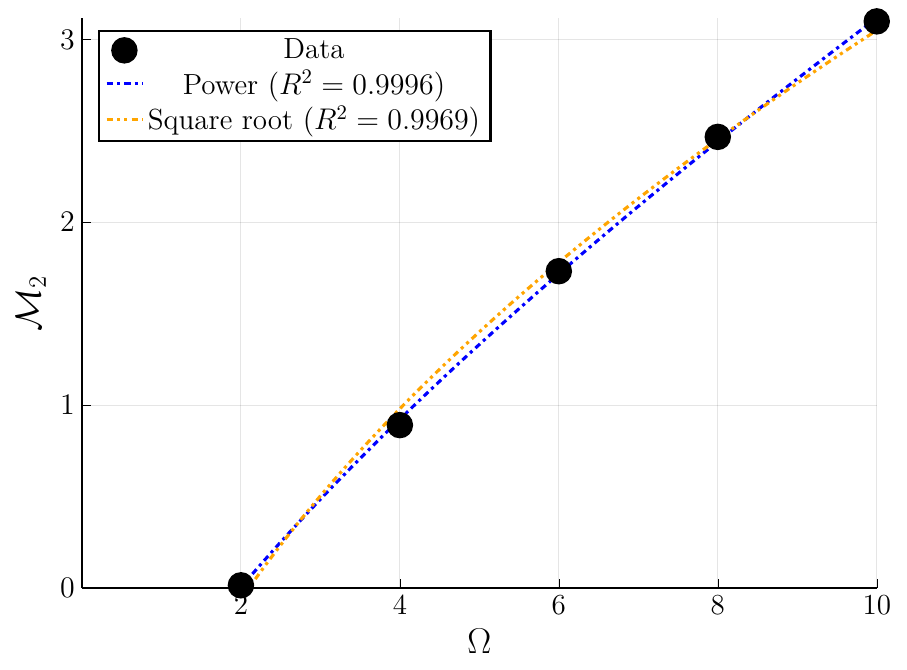}}
     \hfill
     \subfloat[$\vbar=10$, $\gbar=0$]{\includegraphics[width=0.49\textwidth]{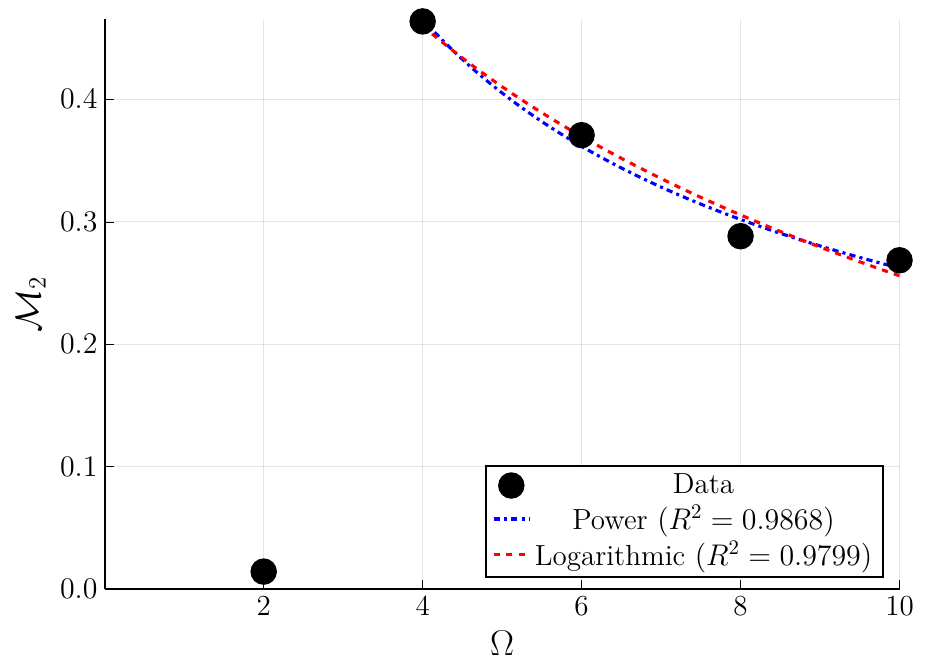}}
\caption{
Scaling of $\M_2$ with particle number. 
In panel d) and e) fits to power laws provide $\M_2 = 1.4126 \times n^{0.7236} -1.4095$ and $\M_2 = 0.7278 \times n^{-0.6017} -0.0142$, respectively.
}
\label{fig:M_2_scaling}
\end{figure}

\begin{figure}[h!]
    \centering
    \subfloat[$\gbar=0$]{\includegraphics[width=0.49\textwidth]{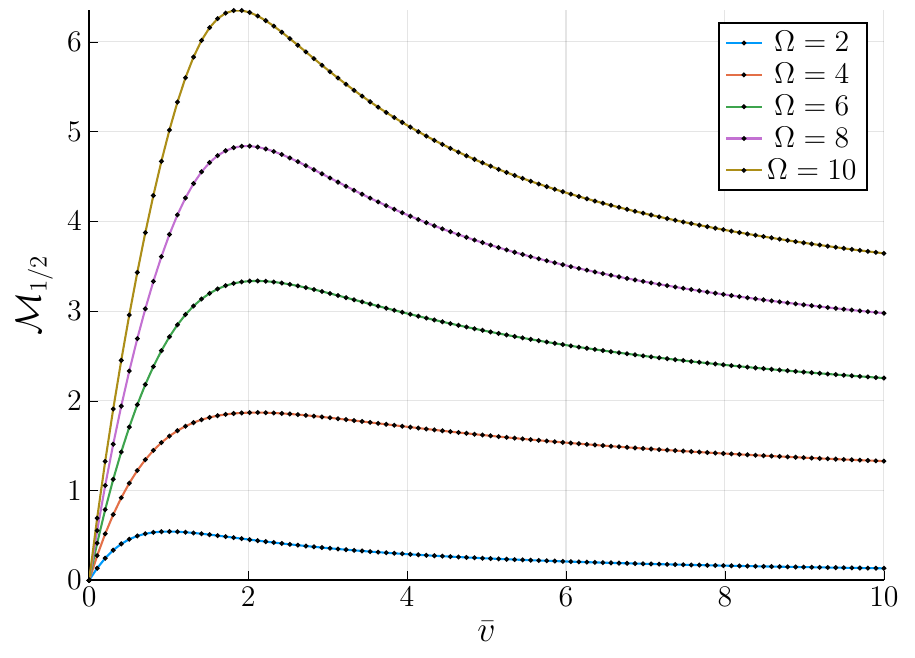}}
     \subfloat[$\vbar=0$]{\includegraphics[width=0.49\textwidth]{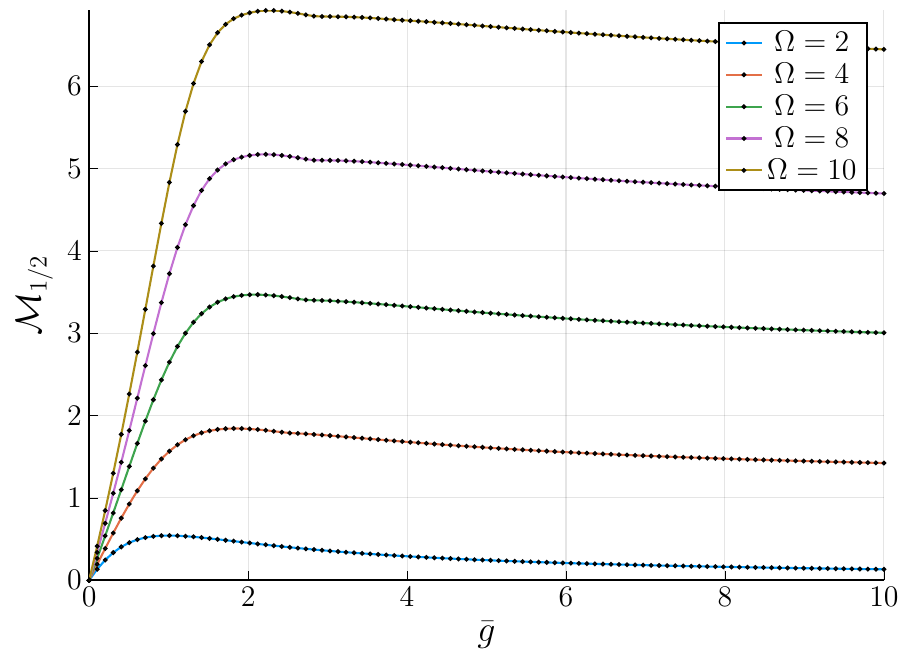}}
     \hfill
     \subfloat[$\M_{1/2}$ along straight line from $(0,0)$ to $(10,9)$.]{\includegraphics[width=0.49\textwidth]{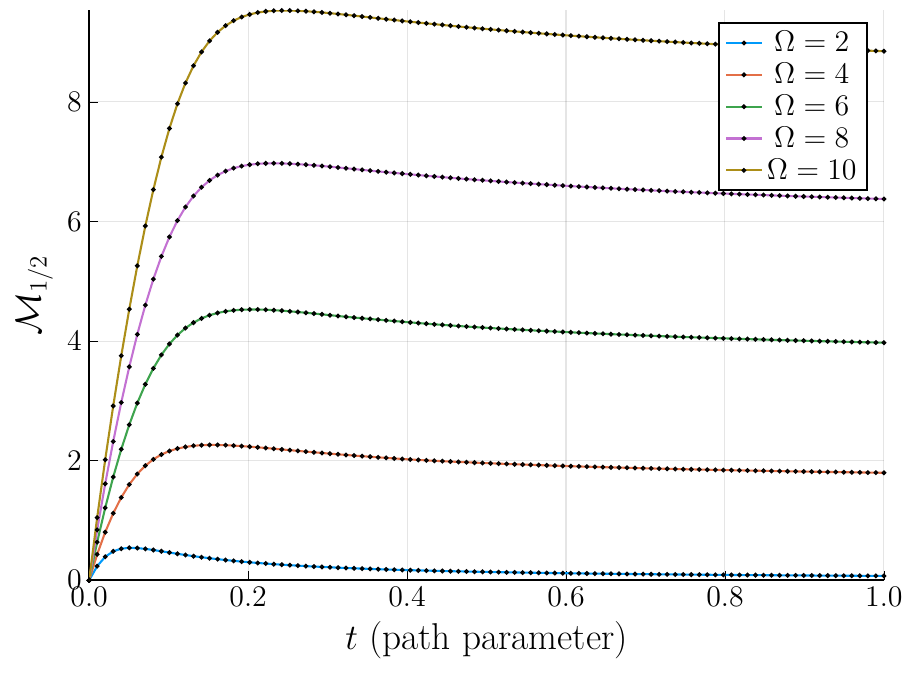}}
     \subfloat[$\vbar=0$, $\gbar=10$]{\includegraphics[width=0.49\textwidth]{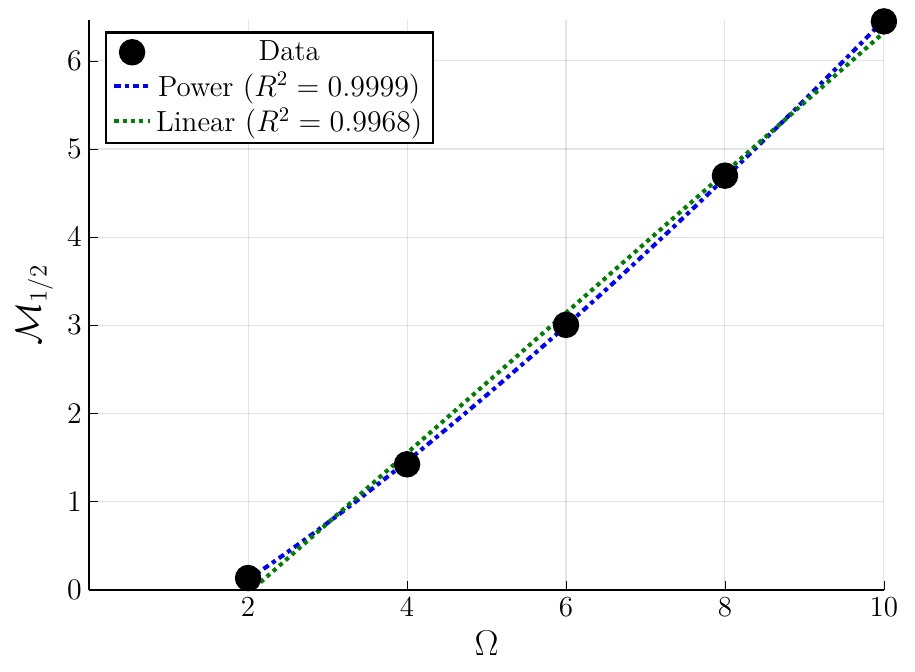}}
     \hfill
     \subfloat[$\vbar=10$, $\gbar=0$]{\includegraphics[width=0.49\textwidth]{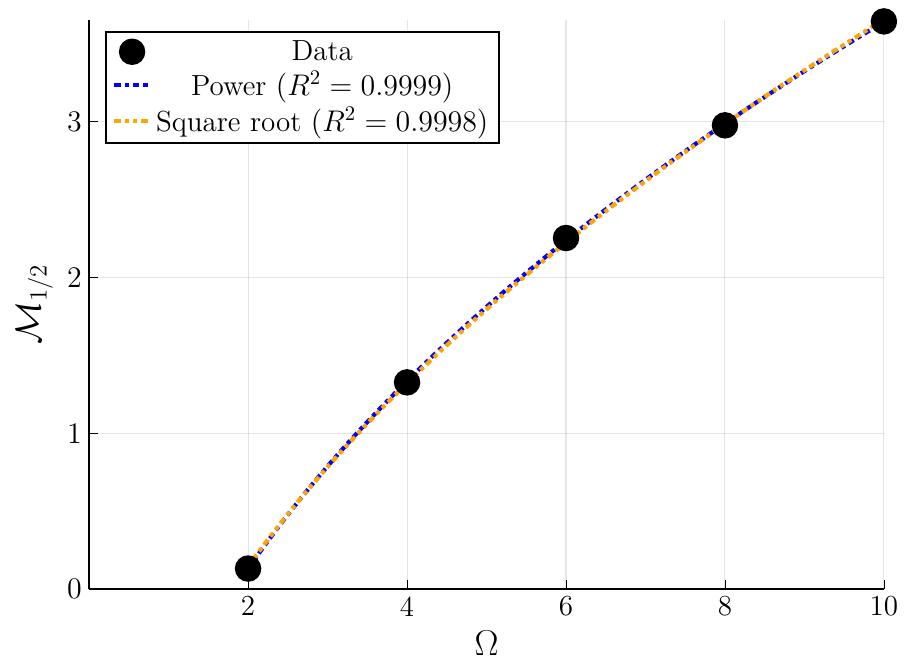}}
\caption{Scaling of $\M_{1/2}$ with particle number.
In panel d) and e) fits to power laws provide $\M_{1/2} = 0.9613 * n^{1.2602} - 0.8433$ and e) Power: $\M_{1/2} = 3.243 * n^{0.4556}-3.1113$, respectively.
}
\label{fig:M_half_scaling}
\end{figure}

%%%%%%%%%%%%%%%%%%%%%%%%%%%%%%%%%%%%%%%%%%%%%%%%%%%%
Figs.~\ref{fig:M_2_scaling} and \ref{fig:M_half_scaling} show more details about the scaling of the SREs $\M_2$ ad $\M_{1/2}$ for different regions of the Agassi phase diagram as $\Omega$ increases.
%

%%%%%%%%%%%%%%%%%%%%%%%%%%%%%%%%%%%%%%%%%%%%%%%%%%%%
\section{Non-Local Non-Stabilizerness}
\label{app:NL_magic}
%%%%%%%%%%%%%%%%%%%%%%%%%%%%%%%%%%%%%%%%%%%%%%%%%%%%
The non-local stabilizer R\'enyi entropy 
\begin{align}
    \M_\alpha^{(n, \, \rm{NL})} (\ket{\Psi}) = \min_{\hat{U}_{A_1} \otimes ... \otimes \hat{U}_{A_n} } \M_\alpha (\hat{U}_{A_1} \otimes ... \otimes \hat{U}_{A_n} \ket{\Psi}) \; ,
\label{eq:NL_NS_app}
\end{align}
where we choose the local degrees of freedom $A_i$ to be mode-pairs (groups of 4 qubits are shown in Fig.~\ref{fig:PAJW_mapping}), as they are are the natural degrees of freedom in the Agassi model.
In the most general case, the $\hat{U}_{A_i}$ unitaries can be up to $16\times 16$ dimensional, which can each be parametrized with $256$ real parameters. In total, this would mean optimizing a set of $256 \times \Omega/2 $ parameters.
Enforcing the relevant symmetries of the model however highly constrains the allowed transformations and significantly reduced the number of parameters to optimize.
First, as the wave function can be taken to be real, due to invariance of the Hamiltonian under complex conjugation, the transformations $\hat{U}_{A_i}$ can be restricted to the orthogonal group which means $120 \times \Omega/2 $ parameters.

Conservation of total particle number can also be exploited. This is written as $[\hat N_{tot},\hat{U}_{tot}]=0$, where 
\begin{align}
    \hat{U}_{tot}=\bigotimes_{i =1}^{n} U_{A_i} \quad \textrm{and} \quad \hat N_{tot} = \sum_{i=1}^n \hat N_{A_i} \, .
\end{align}
Thus the commutator becomes 
\begin{align}
\big[\hat{U}_{tot},\hat N_{tot}\big]
=\left[\bigotimes_i \hat{U}_{A_i},\sum_j \hat N_{A_j}\right]
=\sum_j \left[\bigotimes_i \hat{U}_{A_i},\hat N_{A_j}\right] 
= \sum_j
\left(\bigotimes_{i<j} \hat{U}_{A_i}\right)
\big[\hat{U}_{A_i},\hat N_{A_j} \big]
\left(\bigotimes_{i>j} \hat{U}_{A_i}\right) \; ,
\end{align}
where we used the fact that $\hat N_{A_j}$ acts nontrivially only on mode-pair $A_j$.
The total commutator thus only vanishes if the commutators on every mode-pair vanish, which imposes particle-number conservation in every mode pair. 
The unitary $\hat{U}_{A_i}$ conserves particle number if it is block diagonal in the different particle number sectors. In the Agassi model, since each mode-pair has an even number of fermions, we have three sectors $N_{A_i} = 0,2,4$, spanned by $1,3,1$ states, respectively.
These are $\ket{0000}$ for the $N_{A_i} = 0$ sector, 
$\ket{0101}, \, \ket{1010}, \, (\ket{0110} + \ket{1001})\sqrt{2}$ for the $N_{A_i} = 2$ sector, and
$\ket{1111}$ for the $N_{A_i} = 4$ sector.
Thus, each $U_{A_i}$ can be written as
\begin{align}
    \hat{U}_{A_i} = \mathds{1}_{(N_{A_i}=0)} \oplus \hat{U}_{(N_{A_i}=2)} \oplus \mathds{1}_{(N_{A_i}=4)} \; ,
\end{align}
which acts trivially on the $N_{A_i} = 0,4$ sectors, and $\hat{U}_{N_{A_i}=2} \in SO(3)$, with 3 parameters to optimize. Since the Agassi Hamiltonian is parity symmetric, the unitary $\hat{U}_{A_i}$ is not allowed to mix states with different parities. Since $(\ket{0110} + \ket{1001})\sqrt{2}$ sits in the opposite parity sector from $\ket{0101}$ and $\ket{1010}$, the unitary $\hat{U}_{A_i}$ is not allowed to connect them. Therefore, we have to optimize only one parameter per $\hat{U}_{A_i}$.

Finally, the permutation-symmetry in terms of mode-pairs further implies that the same transformation can be applied on all mode-pairs, {\it i.e} $\hat{U}_{A_i} \equiv \hat{U}$ for all $i$, yielding a constant number of $1$ parameter instead of $1 \times \Omega/2$.

%%%%%%%%%%%%%%%%%%%%%%%%%%%%%%%%%%%%%%%%%%%%%%%%%%%%
\section{Closest Gaussian in Relative Entropy}
\label{app:rel_entropy}
\begin{figure}[ht]
    \centering
    \includegraphics[width=0.5\linewidth]{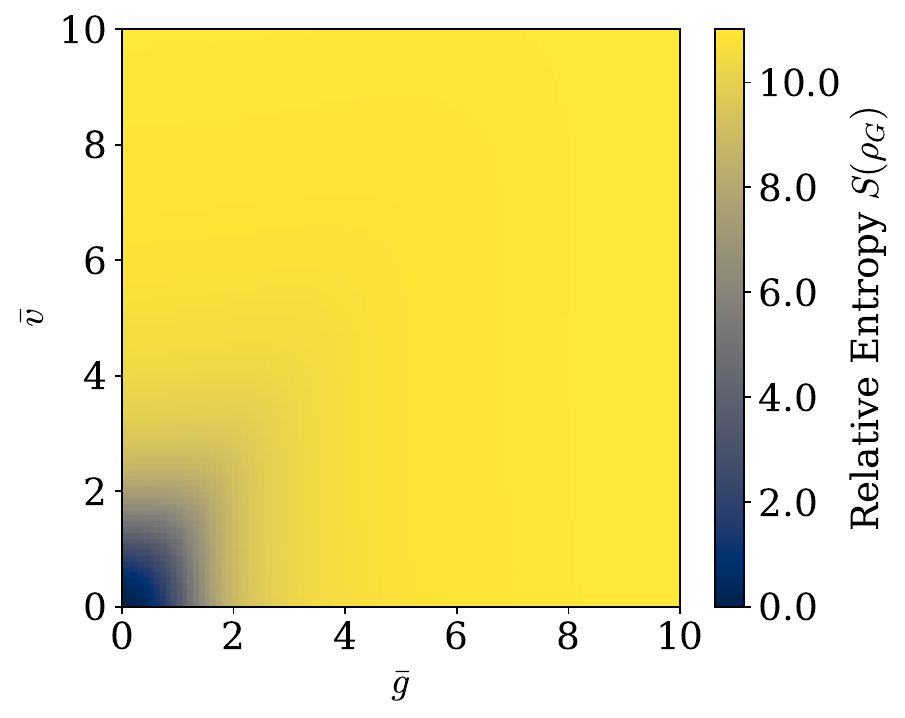}
    \caption{Relative entropy $S(\rho||\rho_G)=S(\rho_G)$ between the Agassi ground state $\rho = \ket{\Psi}\bra{\Psi}$ and the closest Gaussian state $\rho_G$ (with same second order cumulants), for $\Omega=8$ as a function of $\vbar$ and $\gbar$.}
    \label{fig:gaussian_entropy}
\end{figure}

%%%%%%%%%%%%%%%%%%%%%%%%%%%%%%%%%%%%%%%%%%%%%%%%%%%%
Another recently-introduced measure to characterize fermionic magic is the relative entropy 
\begin{align}
    S(\rho||\rho_G) = S(\rho_G)-S(\rho) \; ,
\end{align}
between the state $\rho$ and the Gaussian state $\rho_G$ with same second-order cumulants, {\it i.e.} same entries of the covariance matrix $\Gamma_{ab}$ as $\rho$~\cite{Coffman:2025}. 
Such state $\rho_G$ is the closest Gaussian state in relative entropy~\cite{Coffman:2025}, and $\rho$ differs from $\rho_G$ in higher-order correlations. 
For a pure state $\rho =\ket{\Psi}\bra{\Psi}$, such as the present Agassi ground-state, the relative entropy reduces to $S(\rho||\rho_G)=S(\rho_G)$ which vanishes if and only if $\rho_G$ is a pure Gaussian state. Thus $S(\rho_G)$ provides a measure of non-Gaussian fermionic magic which can be computed via 
\begin{align}
    S(\rho_G) & =-\sum_{m=0}^{2\Omega-1} \left[ \nu_m \log \nu_m +(1-\nu_m)\log(1-\nu_m)\right] \; , 
\end{align}
where $\nu_m=(1+\lambda_m)/2$, and $\lambda_m \in [0,1]$ are the positive eigenvalues of $i\Gamma_{ab}$. 
For the Agassi model, the covariance matrix takes the simple form in Eq.~\eqref{eq:FAF_Agassi}, and thus,
$\lambda_m = |1-2n_m|$, so that,
\begin{align}
    S(\rho_G) & =-\sum_{m=0}^{2\Omega-1} \left[ n_m \log n_m +(1-n_m)\log(1-n_m)\right] \; .
\end{align}
We find that $S(\rho_G)$ exhibits a similar behavior as the FAF, as shown in Fig. \ref{fig:gaussian_entropy}.
%

%%%%%%%%%%%%%%%%%%%%%%%%%%%%%%%%%%%%%%%%%%%%%%%%%%%%
\section{Stabilizer Ground States}
\label{app:stabsforsmall}
%%%%%%%%%%%%%%%%%%%%%%%%%%%%%%%%%%%%%%%%%%%%%%%%%%%%
In this appendix, we detail the procedure for finding stabilizer Hamiltonians in the Agassi model for small systems (small $\Omega$ values). This methodology is then generalized to arbitrary $\Omega$.

\subsection{$\Omega=2$}
For $\Omega=2$ ($N_q=4$ qubits) the Hamiltonian in Eq.~\eqref{eq:QHam} becomes
\begin{align}
\widetilde{H}^{(2)} =& \widetilde{H}_0^{(2)} + \widetilde{H}_{\gbar,diag}^{(2)} +  \widetilde{H}_{\vbar,\gbar}^{(2)} \nonumber \\
=&\frac{1}{4} (-Z_0 + Z_1 -Z_2 + Z_3) 
- \frac{\bar{g}}{4} (2 I + Z_0 + ... + Z_3 +Z_0 Z_2 +Z_1 Z_3)
- (\bar{v} + \bar{g}) \, Q_{0123} \; ,
\label{eq:QHam_2}
\end{align}
where 
\begin{align}
    Q_{0123} = \frac{1}{8} &\bigl( X_0X_1X_2X_3 + X_0X_1Y_2Y_3 - X_0Y_1X_2Y_3 + X_0Y_1Y_2X_3+ Y_0X_1X_2Y_3 - Y_0X_1Y_2X_3 + Y_0Y_1X_2X_3 + Y_0Y_1Y_2Y_3 \bigr) \; .
\end{align}
Thus, the particle-hole and pairing interactions act in the same manner.
We look for stabilizer groups within each Hamiltonian term $\widetilde{H}_0^{(2)}$,  $\widetilde{H}_{\gbar,diag}^{(2)} $, and $  \widetilde{H}_{\vbar,\gbar}^{(2)} $, which dominate the system depending on the regime determined by ($\vbar$, $\gbar$).
The corresponding stabilizer ground-state energies can be easily determined since
for a stabilizer state $\ket{\Psi_s}$, the expectation value of a Pauli string $\hat P$ is given by~\cite{zhu2016clifford}
\begin{align}
    \langle \Psi_s |\hat P| \Psi_s \rangle =
    \begin{cases}
      \pm 1 & \text{if } \hat P \ket{\Psi_s} = \pm \ket{\Psi_s} \; ,\\
      0 & \text{otherwise} \; .\\
    \end{cases}
\end{align}
Thus, the energy of a stabilizer state is simply given by
\begin{equation}
        E\left( \Psi_s  \right) = \langle \Psi_s |\hat H| \Psi_s \rangle = \sum_{P\in \mathcal{G}_N(H)} a_P \underbrace{\langle \Psi_s |\hat P| \Psi_s \rangle}_{0,  \pm 1} \; .
\end{equation}
Since $\ket{\Psi_s}$ is uniquely defined by its stabilizer group $\mathcal{S}$, this equivalently defines the energy of the stabilizer group $\mathcal S$~\cite{Sun:2024lts}.
\begin{itemize}
    \item The stabilizer group $\mathcal{S}_1 = \langle +Z_0, - Z_1, +Z_2, -Z_3 \rangle$ minimizes the energy in $\widetilde{H}_0^{(2)}$. The total stabilizer energy is $E(\mathcal{S}_1)=-(1 + \gbar)$ and the stabilizer state is $\ket{\Psi}_{s,1} = \ket{0101}$, where the qubits are numbered in order $(0)$, $(1)$, $(2)$, $(3)$. This corresponds to the non-interacting ground-state where both particles occupy the lower single-particle states.
    \item $\mathcal{S}_2 = \langle X_0X_1X_2X_3, Y_0Y_1Y_2Y_3, X_0X_1Y_2Y_3, X_0Y_1Y_2X_3 \rangle$ minimizes the energy in $\widetilde{H}_{\vbar,\gbar}^{(2)}$ with energy $E(\mathcal{S}_2)=-\vbar - 2\gbar$ and stabilizer state $\ket{\Psi}_{s,2} = \frac{1}{\sqrt{2}} \left( \ket{0101} + \ket{1010} \right)$. This state is an entangled superposition of the state with two particles on the lower level and the state with two particles on the upper level (Bell state).
    \item Finally, $\mathcal{S}_3 = \langle +Z_0, + Z_1, +Z_2, +Z_3 \rangle$ minimizes the diagonal pairing energy, with $E(\mathcal{S}_3) = - 2 \gbar$ and $\ket{\Psi}_{s,3} = \ket{0000}$, however since this state does not preserve particle number we discard this group in the following.
\end{itemize}

The minimum stabilizer energy $E_{\rm stab} = \min(E(\mathcal{S}_1), E(\mathcal{S}_2))$ which is
\begin{align}
E_{\rm stab} = 
    \begin{cases}
      E(\mathcal{S}_1) = -1 - \gbar & \text{if } \vbar+\gbar > 1 \; ,\\
      E(\mathcal{S}_2) = - \vbar - 2 \gbar & \text{if } \vbar+\gbar > 1 \; .\\
    \end{cases}
\end{align}
The two stabilizer ground states are degenerate at $\vbar+\gbar = 1 $.

\subsection{$\Omega=4$}
Increasing the system size, the particle-hole and pairing interaction play different roles, and the number of Pauli strings in the Hamiltonian increases, which yields more possible stabilizer groups. 
In the case $\Omega=4$ ($N_q =8$), the Hamiltonian in Eq.~\eqref{eq:QHam} reduces to 
\begin{align}
\widetilde{H}^{(4)} =& \widetilde{H}_0^{(4)} + \widetilde{H}_{v}^{(4)} +  \widetilde{H}_{g}^{(4), diag} + \widetilde{H}_{g}^{(4), ndia, I} + \widetilde{H}_{g}^{(4), ndia, II}\; ,
\label{eq:QHam_4}
\end{align}
where 
\begin{align}
\widetilde{H}_0^{(4)} &= \frac{1}{4} (-Z_0 + Z_1 -Z_2 ... +Z_7) \; , \\
\widetilde{H}_{v}^{(4)}  &= -\frac{\vbar}{3} ( Q_{0123} + Q_{0145} + Q_{0167} + Q_{2345}
+ Q_{2367} + Q_{4567} ) \; , \\
\widetilde{H}_{g}^{(4), diag} &= -\frac{\gbar}{12} \big[ 4 I + (Z_0 + Z_1 + ... +Z_7) 
+ Z_0 Z_2 + Z_1 Z_3 + Z_4 Z_6 + Z_5 Z_7 \big] \; , \\
\widetilde{H}_{g}^{(4), ndia, I} &= -\frac{\gbar}{3} (Q_{0123} + Q_{4567}) \; \\
\widetilde{H}_{g}^{(4), ndia, II} &= -\frac{\gbar}{3} (\overline{Q}_{012,456} + \overline{Q}_{012,567} + \overline{Q}_{123,456} + \overline{Q}_{123,567}) \; . 
\end{align}
One can find several stabilizer groups from the various Hamiltonian terms in Eq.~\eqref{eq:QHam_4}:
\begin{itemize}
    \item Terms from $\widetilde{H}_0$: $\mathcal{S}_1 = \langle +Z_0, - Z_1, +Z_2, -Z_3 .... -Z_7 \rangle$ with total stabilizer energy $E(\mathcal{S}_1) = -2 - \frac{2 \gbar}{3}$ and $\ket{\Psi}_{s,1} = \ket{01010101}$. This again corresponds to the non-interacting ground state where all particles occupy the lower single-particle states.
    \item Terms from $\widetilde{H}_{v}$: $\mathcal{S}_2 = \langle X_0X_1X_2X_3, Y_0Y_1Y_2Y_3, X_0X_1Y_2Y_3, X_0X_1X_4X_5, X_0X_1Y_4Y_5, X_0X_1X_6X_7, X_0X_1Y_6Y_7,$ $ X_0 Y_1 Y_2 X_3 X_4 Y_5 Y_6 X_7\rangle$ with energy $E(\mathcal{S}_2)  = - \frac{2\gbar}{3}  - \vbar$ and 
    \begin{align}
        \ket{\Psi}_{s,2} = -\frac{1}{\sqrt{8}} &\big( |01010101\rangle + |01011010\rangle  +|01100110\rangle + |01101001\rangle \nonumber \\
        &+ |10010110\rangle +|10011001\rangle + |10100101\rangle + |10101010\rangle \big) \; .
    \end{align}
    \item Terms common to $\widetilde{H}_{v}$ and $\widetilde{H}_{g}^{ndiag,I}$: $\mathcal{S}_3 = \langle X_0X_1X_2X_3, Y_0Y_1Y_2Y_3, X_0X_1Y_2Y_3, X_0Y_1Y_2X_3, X_4X_5X_6X_7,$ $ Y_4Y_5Y_6Y_7, X_4X_5Y_6Y_7, X_4Y_5Y_6X_7 \rangle$ with $E(\mathcal{S}_3)  = - \frac{2}{3} (\vbar + 2\gbar)$.
    The corresponding stabilizer state is a tensor product of two entangled mode-pairs:
    \begin{align}
        \ket{\Psi}_{s,3} = - \left( \frac{\ket{0101} + \ket{ 1010}}{\sqrt{2}} \right)^{\otimes{2}}
    \end{align}
    \item Terms from $\hat{H}_{g}^{ndiag,II}$: For example, stabilizer group $\mathcal{S}_4^{(a)} = \langle s_{012,456}, s_{123,567} \rangle$, where 
    \begin{align}
        s_{ijk,lmn} \equiv (X_i Z_j X_k X_l Z_m X_n, Y_i Z_j Y_k Y_l Z_m Y_n, 
        -X_i Z_j X_k Y_l Z_m Y_n, X_i Z_j Y_k Y_l Z_m X_n) \; ,
    \end{align} 
    with $E(\mathcal{S}_4^{(a)} ) = - \frac{4\gbar}{3}$ and 
    \begin{align}
        \ket{\Psi}_{s,4a} =\frac{1}{2} ( |00001111\rangle - |01011010\rangle
        - |10100101\rangle + |11110000\rangle ) \; ,
    \end{align}
    and, stabilizer group $\mathcal{S}_4^{(b)} = \langle s_{012,567}, s_{123,456} \rangle$ with the same total stabilizer energy $E(\mathcal{S}_4^{(b)} ) =E(\mathcal{S}_4^{(a)} )$ and 
    \begin{align}
        \ket{\Psi}_{s,4b} =\frac{1}{2} ( |00001111 \rangle - |01010101\rangle
        - |10101010\rangle + |11110000\rangle) \; .
    \end{align}
    \item Terms from $\hat{H}_{g}^{diag}$: $\mathcal{S}_5 = \langle +Z_0, + Z_1, +Z_2, +Z_3, ... +Z_7 \rangle$ with energy $E(\mathcal{S}_5)  = - \frac{4\gbar}{3}$ and $\ket{\Psi}_{s,5} = \ket{00000000}$, which again yields a stabilizer state $\ket{\Psi}_{s,5}$ which violates particle-number symmetry, and thus is not retained.
\end{itemize}

%%%%%%%%%%%%%%%%%%%%%%%%%%%%%%%%%%%%%%%%%%%%%%%%%%%%
\section{Stabilizer expansions, truncations and rotations}
\label{app:LRS_trunc}
%%%%%%%%%%%%%%%%%%%%%%%%%%%%%%%%%%%%%%%%%%%%%%%%%%%%
In this appendix we study the convergence of truncated stabilizer expansions 
\begin{align}
    \ket{\Psi}_{r} &= \sum_{i=1}^r  c_i \ket{\Psi}_{s,i} \nonumber \\
    & = c_1 \ket{\Psi}_{s,1} + c_2 \ket{\Psi}_{s,2} + c_3 \ket{\Psi}_{s,3} 
    + \sum_{a} c_{4,a} \ket{\Psi}_{s,4}^{(a)} \; ,
\label{eq:LRS_app}
\end{align}
in different regimes of the Agassi Hamiltonian.

\begin{figure}[h]
    \centering
    \subfloat[$\vbar=0.0,\gbar=2.0$]{\includegraphics[width=0.45\textwidth]{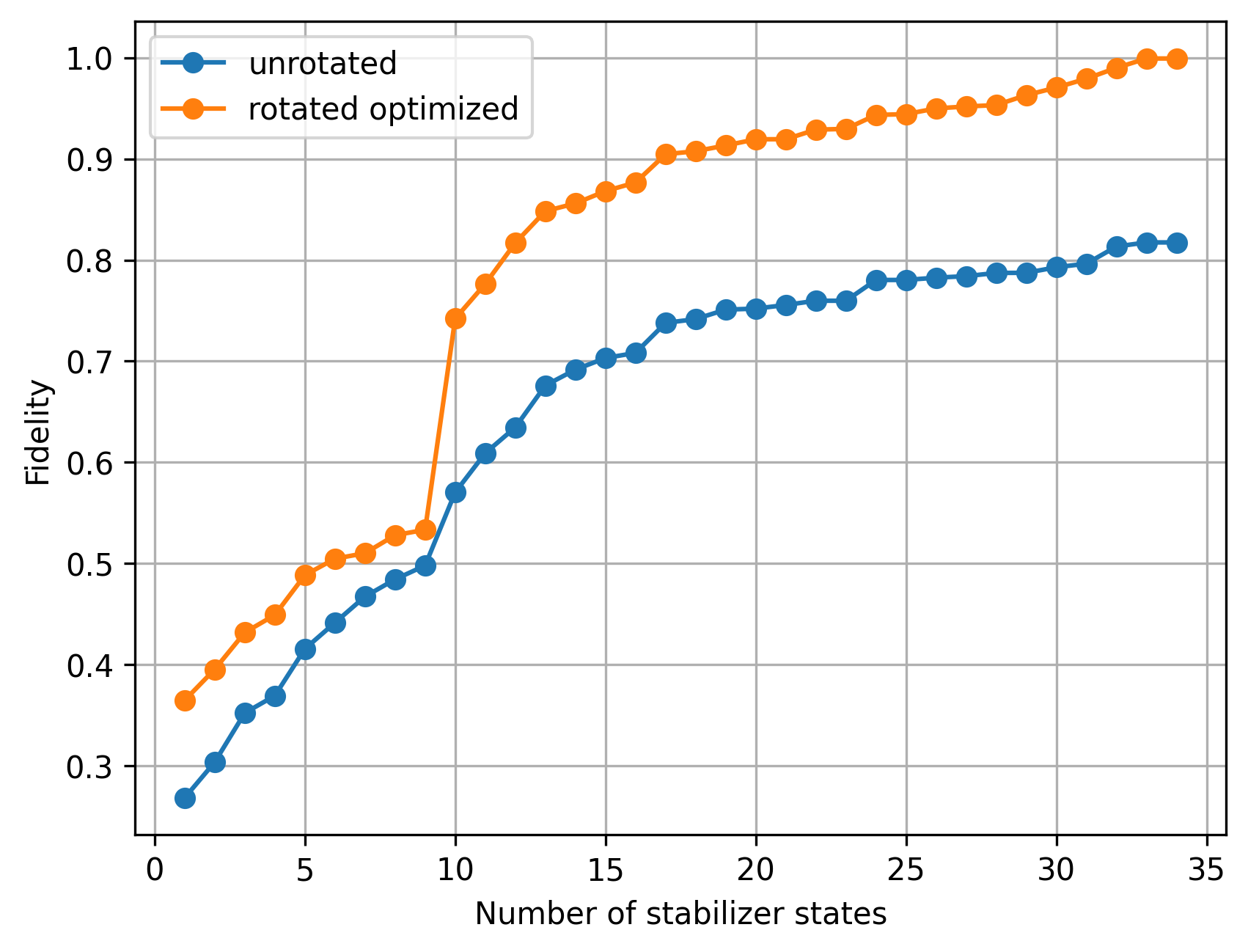}}
    \subfloat[$\vbar=2.0,\gbar=0.0$]{\includegraphics[width=0.45\textwidth]{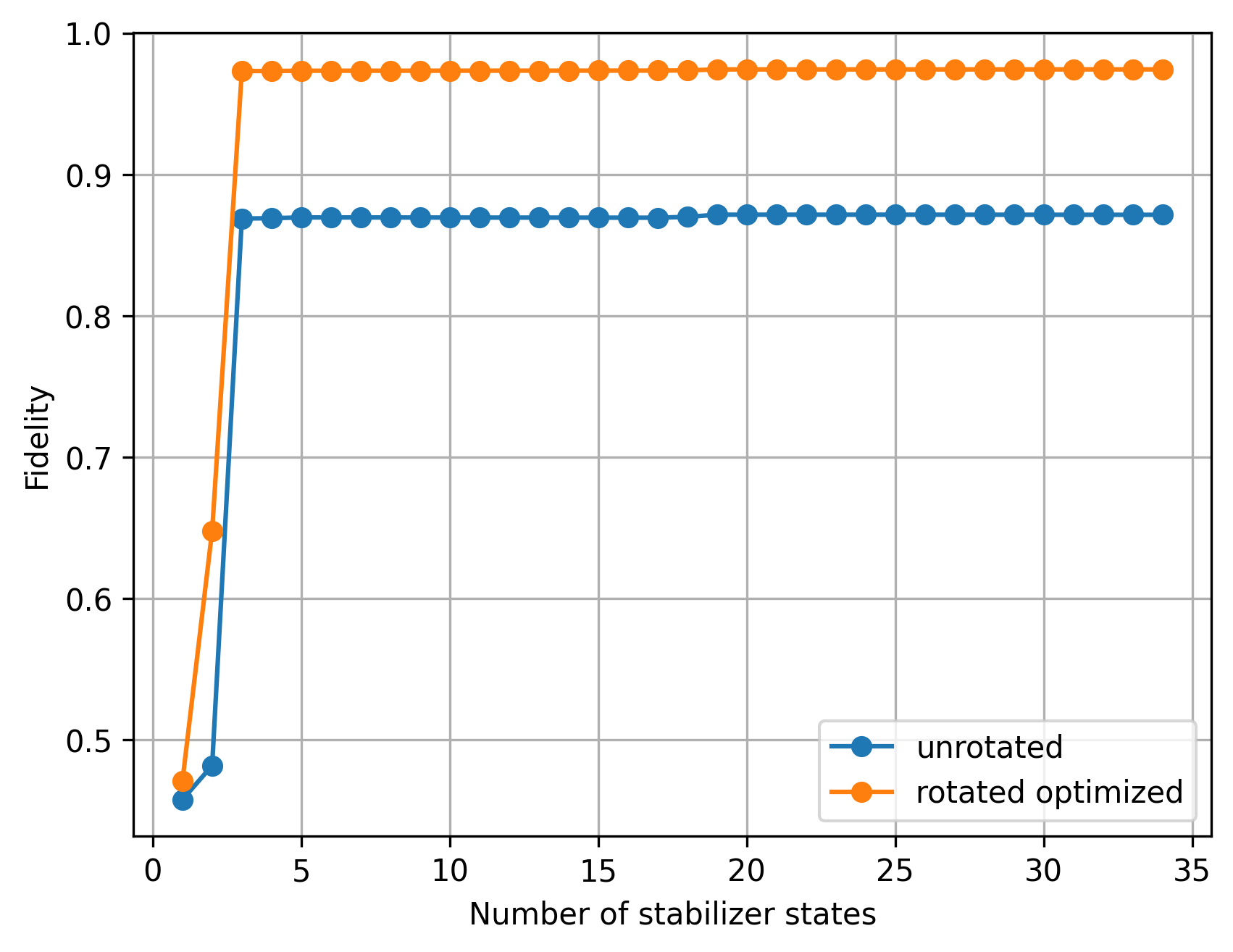}}
    %\subfloat[]{\includegraphics[width=0.3\textwidth]{}}
    \caption{Convergence of the fidelity of the stabilizer-state expansion in Eq.~\eqref{eq:LRS_app} for various parameter values $(\vbar,\gbar)$. Note that the order of stabilizer state is fixed.}
    \label{fig:trunc_LRS}
\end{figure}

Results are shown in Fig.~\ref{fig:trunc_LRS} for a system with $\Omega=8$, for which the maximum number of stabilizer states included in Eq.~\eqref{eq:LRS_app} (rank) is $r=34$.
For each $(\vbar, \gbar)$ value we show results obtained with stabilizer states $\ket{\Psi}_{s,i}$ built on the "bare" basis, and those obtained with stabilizer states $\ket{\Psi'}_{s,i} = \hat{U} \ket{\Psi}_{s,i}$ built on the rotated mode-pair basis, where the symmetry-preserving transformation $\hat{U} = \otimes_{i=1}^{\Omega/2} \hat{U}_{A_i}$ is optimized to minimize energy.
In the regime near the normal-superfluid phase transition (left panel), it is seen that a wavefunction with $\approx 90\%$ fidelity can be obtained with about $17$ rotated stabilizer states, while the non-rotated version only allows for $\approx 82 \%$ fidelity with twice the number of stabilizer states.

\end{document}